\documentclass[a4paper,amsmath,amssymb,pre,twocolumn,superscriptaddress]{revtex4-1}

\usepackage{amsmath}
\usepackage{caption, subcaption}
\usepackage{graphicx}
\usepackage{epstopdf}
\usepackage{hyperref}
\usepackage{amsthm}
\let\label\relax

\def\er{Erd\H{o}s-R\'enyi~}
\newtheorem{theorem}{Theorem}[section]
\newtheorem{lemma}[theorem]{Lemma}

\theoremstyle{remark}

\begin{document}

\title{The limitations of discrete-time approaches to continuous-time contagion dynamics}

\author{Peter G. Fennell}
\affiliation{MACSI, Department of Mathematics and Statistics, University of Limerick, Ireland}
\author{Sergey Melnik}
\affiliation{MACSI, Department of Mathematics and Statistics, University of Limerick, Ireland}
\author{James P.~Gleeson}
\affiliation{MACSI, Department of Mathematics and Statistics, University of Limerick, Ireland}

\begin{abstract}
{Continuous-time Markov process models of contagions are widely studied, not least because of their utility in predicting the evolution of real-world contagions and in formulating control measures. It is often the case, however, that discrete-time approaches are employed to analyze such models or to simulate them numerically. In such cases, time is discretized into uniform steps and transition rates between states are replaced by transition probabilities. In this paper, we illustrate potential limitations to this approach. We show how discretizing time leads to a restriction on the values of the model parameters that can accurately be studied. We examine numerical simulation schemes employed in the literature, showing how synchronous-type updating schemes can bias discrete-time formalisms when compared against continuous-time formalisms. Event-based simulations, such as the Gillespie algorithm, are proposed as optimal simulation schemes both in terms of replicating the continuous-time process and computational speed. Finally, we show how discretizing time can affect the value of the epidemic threshold for large values of the infection rate and the recovery rate, even if the ratio between the former and the latter is small.}
\end{abstract}

\maketitle

\section{Introduction}

A feature of our environment is the existence of networks, from real-life human contact networks, to virtual networks such as online social networks, to functional and technological networks such as transport networks and the Internet~\cite{newman2010networks}. Networks form a medium for contagions, which spread from node to node through the links of the networks. Contagions can be physical~\cite{pastor2001epidemic,anderson1979population}, cultural~\cite{axelrod1997dissemination,watts2002simple}, societal~\cite{o2015mathematical,centola2010spread,hodas2014simple}, financial~\cite{gleeson2012systemic,gai2010contagion,may2010systemic} and the modelling of such contagions~\cite{gleeson2014simple,melnik2013,gleeson2014competition,gleeson2011high,gleeson2013} -- along with the understanding of the suitability of various modelling approaches~\cite{gleeson2012accuracy,melnik2011unreasonable,faqeeh2015network} -- is vital for matters of the upmost public importance~\cite{gomes2014assessing,gonzalez2011dynamics,lazer2009life}.

A common modelling paradigm for studying contagions is the framework of continuous-time Markov processes~\cite{van2014performance,barrat2008dynamical,cox1977theory}, where events (such as the infection of a susceptible individual by an infected individual) occur at certain \emph{rates}. The most well known of these models are epidemiological compartment models~\cite{anderson1979population} which, although introduced as models of disease spread~\cite{kermack1927contribution}, are also widely used as models of social contagions such as the diffusion of information and innovations~\cite{toole2012modeling,kitsak2010identification,daley1965stochastic}. Continuous-time Markov process models can provide valuable insights into contagion processes, and have real value in both predicting and controlling contagious outbreaks~\cite{ferguson2006strategies,degli2008mitigation,balcan2010modeling}.

One avenue to study continuous-time Markov process models is by using discrete-time approximations~\cite{allen1994some,wang2003epidemic,zou2007modeling,chakrabarti2008epidemic,gomez2010discrete,balcan2010modeling,de2013anatomy,wei2013competing,granell2013dynamical,cozzo2013contact,valdano2015analytical,gomez2016explosive}. Such approaches can be either numerical (i.e., synchronous updating Monte Carlo simulations) or theoretical. In a discrete-time approach, time is discretized into time-steps of length $\Delta t$ (which usually takes the value $\Delta t=1$), and events occur with certain \emph{probabilities}. These probabilities are known as the state transition probabilities, and are simply the product of the corresponding rate and the time-step $\Delta t$.

Although discrete-time approaches correspond to their continuous-time counterpart in the limit $\Delta t \rightarrow 0$, they can significantly differ in the case that $\Delta t$ is finite. Allen, in her work~\cite{allen1994some}, shows that discrete-time susceptible-infected-susceptible (SIS) and susceptible-infected-recovered (SIR) models can produce complex behaviour such as period doubling and chaotic effects for sufficiently large values of the time step and/or contact rate. This behaviour is not possible in the continuous-time SIS and SIR models, and is thus no more than an artefact of discretizing time. Similarly, Gomez \emph{et al.}~\cite{gomez2011nonperturbative} observe that differences between continuous and discrete-time SIS dynamics are ``substantial'' when an arbitrary time step of $\Delta t=1$ is employed. An understanding of the discrepancies introduced as a result of discretizing time is thus important, allowing us to gauge the validity of discrete-time approaches and when they may accurately be employed.

In this paper, we show the limitations of discrete-time approaches when used to study continuous-time contagion dynamics. Our message is clear -- that the accuracy of such methods will be poor if state transition probabilites are too large, leading to deviations from the underlying continuous-time process. The repercussions of this are manifold. Discrete-time theoretical approaches can be significantly inaccurate for large values of the contagion parameter values (such as infection and recovery rates), and thus the analysis of such approaches will not be valid. Furthermore, discrete-time Monte Carlo simulations -- often used as a gold standard~\cite{pastor2001epidemic,barthelemy2004velocity,wang2003epidemic} -- can be inaccurate for large parameter values, and such inaccurate simulations can lead to misleading conclusions. We illustrate this latter point with an example from the literature in Section 4. Our results shed light on the inaccuracies caused when the state transition probabilities of discrete-time approaches are too large, and provide guidelines for the appropriate use of such approaches.

\section{Continuous-time contagion dynamics and the discrete-time approximation}
\label{sec:SIS}

To begin, we describe in some detail both continuous and discrete-time Markov processes to illustrate mathematically the difference between the two. In continuous-time Markov processes, events are described by rates $\lambda$, while events in the discrete-time analogue are described by transition probabilities $\tilde{\lambda}$, where $\tilde{\lambda} = \lambda \Delta t$. In the course of our analysis we focus on the specific example of SIS dynamics; however, our analysis holds for any continuous-time Markovian dynamics, with the core message being the limitations on the size of the transition probabilites $\tilde{\lambda}$ for which discrete-time approaches are accurate.

\subsection{Continuous-time SIS dynamics}
\label{sec:discrete_SIS}

Consider SIS dynamics taking place on a network of $N$ nodes. This is a continuous-time Markov process where at any time $t$ each node $i$ in the network has a corresponding state $X^i_t$ which is either susceptible ($X^i_t=S$) or infected ($X^i_t=I$)~\cite{bailey1975mathematical, liggett1999, van2014performance}. The states of each node in the network change dynamically over time. Susceptible nodes become infected through each of their infected neighbours at a rate $\beta$ per infected neighbour, while infected nodes recover at a rate $\mu$. ``Rate'' here refers to instantaneous transition rates, which in continuous-time dynamics define the transitions between states; these are defined in terms of probabilities as
\begin{align}
    \mu &= \lim_{\Delta t \rightarrow 0}\frac{P(X^i_{t+\Delta t} = S|X^i_t = I)}{\Delta t}, \label{eq:transitionrateM} \\
    \beta &= \lim_{\Delta t \rightarrow 0}\frac{P(X^i_{t+\Delta t}=I\mbox{ via }j|X^i_t = S, X^j_t = I)}{\Delta t}
    \label{eq:transitionrateB}
\end{align}
where \{$X^i_{t+\Delta t}=I\mbox{ via }j$\} is the event that a susceptible node $i$ became infected through an infected neighbour $j$. The fraction terms in the right hand sides of Eqs.~\eqref{eq:transitionrateM} and \eqref{eq:transitionrateB} are the probabilities of state changes per unit time and taking the limit of these
fractions as $\Delta t \rightarrow 0$ leads to the concept of transition rates. In general, we can define $r_i$ as the rate at which a node $i$ changes from its current state to the opposite state; this is given by
\begin{equation}
  r_i =
    \begin{cases}
        \beta m_{i,t} & \mbox{if } X^i_t=S \\
        \mu & \mbox{if }X^i_t=I
    \end{cases},
  \label{eq:transitionrateSIS}
\end{equation}
 where $m_{i,t}$ is the number of infected neighbours of node $i$ at time $t$.

\begin{figure*}
  \centering
  \includegraphics[width=0.75\textwidth]{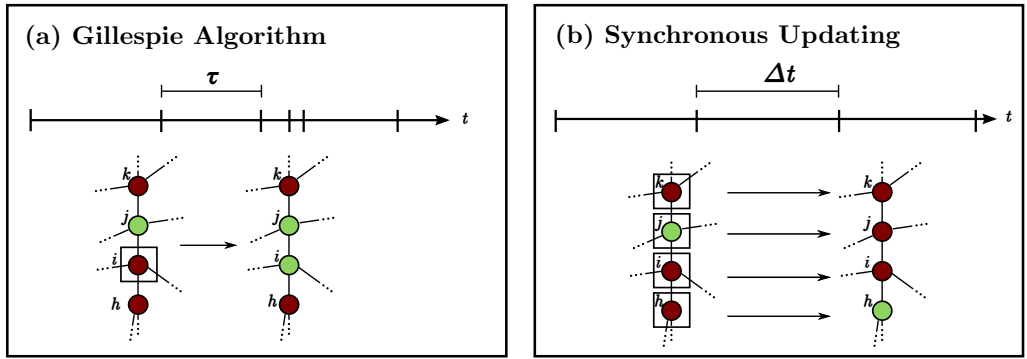}
  \caption{Schematics of both (a) the Gillespie algorithm and (b) the synchronous updating scheme. Vertical ticks on the $t$-axis indicate the moments through which the simulation advances --- in synchronous updating the interval between these moments is a fixed time-step with value $\Delta t$ while in the Gillespie algorithm the interval is a random variable $\tau$ given by Lemma~\ref{lemma:1}. The light green and dark red circles are nodes in the network which are in the susceptible and infected states respectively. A square around a node means that the node has been chosen for updating at a certain moment and may change its state. In the Gillespie algorithm a node is chosen according to Lemma~\ref{lemma:2} and will always change its state while in the synchronous updating scheme every node has the chance to change state and will do so with a probability that depends on their state and the states of their nearest neighbours.}
  \label{fig:schematic}
\end{figure*}

The evolution of the dynamics in the network can be fully described by the master equation for the Markov process~\cite{gardiner1985handbook, barrat2008dynamical}. If we denote by $\textbf{Y}_t = \{X^i_t\}_{i=1}^N$ the state of the network at time $t$, and by $p(\textbf{y},t)$ the probability that the network is in state $\textbf{Y}_t=\textbf{y}$, then the master equation is given by
\begin{equation}
    \frac{d}{dt}p(\textbf{y},t) = \sum_{\textbf{y}' \neq \textbf{y}}\Big(p(\textbf{y}',t)r_{\textbf{y}' \rightarrow \textbf{y}} - p(\textbf{y},t)r_{\textbf{y} \rightarrow \textbf{y}'} \Big), 
    \label{eq:masterequation}
\end{equation}
with initial conditions $p(\textbf{y},0) = p_0(\textbf{y})$. Here $r_{\textbf{y} \rightarrow \textbf{y}'}$ is the instantaneous rate at which the network changes from state $\textbf{y}$ to $\textbf{y}'$ and is fully determined by the network structure and the transition rates $\mu$ and $\beta$. While the master equation is the gold standard -- exactly describing the evolution of SIS dynamics -- the dimension of its sample space $\Omega$ is $2^{N}$, which in general is prohibitively large for analytical or numerical studies. One way to tackle this problem is to study the dynamics as a series of individual transitions between states. In continuous-time dynamics nodes change state one at a time, or asynchronously~\cite{liggett1985particle}. Given the state of the network, the probability distributions governing both the length of time until the next state change and the node which will change state can be constructed. These are given by the following lemmas (of which rigorous derivations can be found in the literature, e.g., \cite[Chapter~10]{van2014performance}):

\begin{lemma}
    Let $\tau$ be the \emph{holding time} of the network, the length of time that the network remains in its current state before changing to the next state. Then $\tau$ is an Exponentially distributed random variable and the parameter of the distribution is the sum of the individual node transition rates, i.e., $\sum_{i=1}^{N}r_i$.
    \label{lemma:1}
\end{lemma}
\begin{lemma}
    The probability that the next node in the network to change state will be node $i$ is $r_i/\sum_{j=1}^Nr_j$.
    \label{lemma:2}
\end{lemma}
Lemmas~\ref{lemma:1} and \ref{lemma:2} describe how the network probabilistically evolves from one state to another. They are the basis of continuous-time stochastic simulation methods such as the well-known Gillespie algorithm, also known as the Stochastic Simulation algorithm or Kinetic Monte Carlo~\cite{gillespie1977, ferreira2012epidemic, vestergaard2015temporal, voter2007introduction}. Such simulations are often referred to as event-based simulations because the time intervals are not fixed but rather correspond to the time between consecutive state-changes in the system. At each step in such algorithms, time advances by an amount $\tau$ and node $i$ changes its state, where $\tau$ and $i$ are random numbers drawn according to Lemmas~\ref{lemma:1} and \ref{lemma:2} (Fig.~\ref{fig:schematic}). Stochastic simulations give the opportunity to construct $p(\textbf{y},t)$ empirically by running multiple realizations of the stochastic process and aggregating over an ensemble of realizations. Such simulations are statistically exact as they are fully based on Lemmas~\ref{lemma:1} and \ref{lemma:2} which are derived without approximation from the axioms of the Markov process.

\subsection{The discrete-time approach}
\label{sec:discrete_SIS}

In a discrete-time framework, time is no longer treated as a continuous variable but rather takes the form of a discrete variable which advances in time intervals of length $\Delta t$. Instantaneous transition rates are replaced by transition probabilities. In a single time interval, susceptible nodes become infected through their infected neighbours with probability $\tilde{\beta} = \beta\Delta t$ per infected neighbour, while infected nodes recover with probability $\tilde{\mu} = \mu\Delta t$. Note that $\Delta t$ is often assumed to take the value $\Delta t=1$, but even in this case it should be included in the expression for $\tilde{\beta}$ and $\tilde{\mu}$ to clarify that a rate needs to be multiplied by a time step before it can be expressed as a probability.

The discretization of time in this manner leads to two deviations from the continuous-time process. These deviations arise through both the transition probabilities which are used in place of transition rates as well as the parallel (synchronous) state-changes in discrete-time systems that are uncharacteristic of continuous-time dynamics. To understand the roots of the deviations introduced through the transition probabilities we can examine the definitions of $\mu$ and $\beta$ as rates given in Eqs.~\eqref{eq:transitionrateM} and \eqref{eq:transitionrateB}. These equations can be re-arranged to give transition probabilities in terms of these rates, i.e.,
\begin{align}
    &P(X^i_{t+\Delta t} = S|X^i_t = I) = \mu \Delta t \label{eq:transitionprobM} \\
    &P(X^i_{t+\Delta t}=I\mbox{ via }j|X^i_t = S, X^j_t = I) = \beta \Delta t
    \label{eq:transitionprobB}
\end{align}
where in this case $\Delta t$ is an infinitesimally small length of time. In the case that $\Delta t$ is not infinitesimally small, Eqs.~\eqref{eq:transitionprobM} and \eqref{eq:transitionprobB} become approximations. In a time interval of length $\Delta t$ in the continuous-time Markov process, the exact probability that an infected node will recover is $1-e^{-\mu\Delta t}$ while the probability that a susceptible node will become infected by a given infected neighbour is $1-e^{-\beta\Delta t}$. The transition probabilities $\tilde{\mu}$ and $\tilde{\beta}$, the right-hand side of Eqs.~\eqref{eq:transitionprobM} and \eqref{eq:transitionprobB}, are approximations to $1-e^{-\mu\Delta t}$ and $1-e^{-\beta\Delta t}$ respectively and an important question then arises of the effect on the dynamics to which these approximations have.

\begin{figure}[!t]
  \centering
  \includegraphics{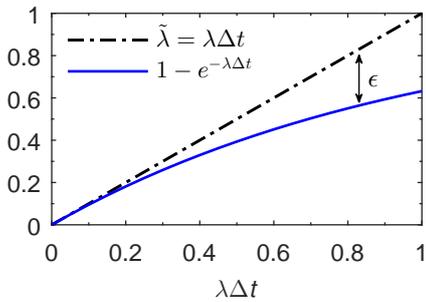}
  \caption{The actual probability (blue solid line) that a rate $\lambda$ event will occur in a time
    step of length $\Delta t$ plotted along with the approximate probability $\tilde{\lambda}$
    (black dash-dotted line) as used in discrete-time formalisms. The error $\epsilon$ is defined as the absolute distance between the two.}
  \label{fig:cdtapprox}
\end{figure}

Figure~\ref{fig:cdtapprox} shows the actual probability $1-e^{-\lambda\Delta t}$ along with the discrete-time probability $\tilde{\lambda} = \lambda\Delta t$, where we use the parameter $\lambda$ to represent either $\mu$ or $\beta$. We also plot the error $\epsilon$ which is defined as the difference between the discrete-time probability and the actual probability. When $\tilde{\lambda} < 0.1$, $\epsilon < 0.01$ and so the approximation is
fairly accurate in this range. For larger values of the state transition probability $\tilde{\lambda}$,
however, the approximation differs significantly from the true values. At $\tilde{\lambda} =
0.5$, $\epsilon \approx 0.1$ and  when $\tilde{\lambda} = 1$, $\epsilon \approx 0.37$. These individual errors can accumulate and have significant implications on the dynamics as a whole; indeed we show empirically in Sections~\ref{sec:comparison} and \ref{sec:continvsdiscrete} that although discrete-time approaches can be very accurate when $\tilde{\mu}$ and $\tilde{\beta}$ are very small they begin to lose accuracy when $\tilde{\mu}$ and $\tilde{\beta}$ are of the order of magnitude of $10^{-1}$.

Secondly, we comment on the synchronous updating nature of discrete-time approaches. This is in contrast to the continuous-time process where nodes change state asynchronously and the change of state of one node immediately affects the transition rates of the other nodes (Fig.~\ref{fig:schematic}). The strength of effect will depend on the transition probabilities, as the values $\tilde{\mu}$ and $\tilde{\beta}$ dictate the number of state changes that take place in each time step and thus the propensity of multiple nodes to change state at the same time.

Thus we arrive at a simple conclusion: the values of $\tilde{\mu}$ and $\tilde{\beta}$ (and thus $\mu$, $\beta$ and $\Delta t$) used in discrete-time approaches should be controlled so that these approaches are accurate representations of the continuous-time process. For large values of $\mu$ or $\beta$, the time step $\Delta t$ should be small while if $\Delta t = 1$, as in the case of the majority of discrete-time approaches, the values of $\mu$ and $\beta$ should be relatively small. Throughout the rest of this paper we will give empirical evidence of this conclusion.

Finally, we comment on discrete-time numerical simulation schemes that are used to stochastically simulate SIS dynamics. A commonly used simulations scheme is synchronous updating, also referred to as Rejection sampling (Fig.~\ref{fig:schematic}) \cite{wang2003epidemic, pastor2001epidemic, vestergaard2015temporal, neal2003slice}. In this case, time advances in steps of one time unit, i.e., $\Delta t=1$. In a single time unit, a susceptible node will become infected by its infected neighbours with probability $\tilde{\beta}$ per infected neighbour while infected nodes become susceptible with probability $\tilde{\mu}$. Synchronous updating simulations are statistically exact realizations of the discrete-time dynamics; these dynamics are fully described by the discrete-time master equation
\begin{equation}
    p(\textbf{y},t+1) = \sum_{\textbf{y}'}p(\textbf{y}',t)q_{\textbf{y}'\rightarrow \textbf{y}},
\end{equation}
where $q_{\textbf{y}'\rightarrow \textbf{y}}$ is the probability that the network changes from state $\textbf{y}'$ to state $\textbf{y}$ in a time-step of length $\Delta t = 1$ and is fully determined by the network structure and the transition probabilities $\tilde{\mu}$ and $\tilde{\beta}$~\cite{barrat2008dynamical}. Because synchronous updating simulations exactly mimic the discrete-time dynamics and master equation they will be used throughout this paper to gauge the accuracy of the discrete-time approach.

In the remainder of the paper, we show how the approximations introduced in discrete-time approaches can lead to misrepresentation of the actual continuous-time dynamics. We begin in the next section by examining the discrete-time approximations of Eqs.~\eqref{eq:transitionprobM} and \eqref{eq:transitionprobB} for fixed $\mu$ and $\beta$ and various values of $\Delta t$. We show that discrete-time dynamics can accurately reproduce continuous-time dynamics for small values of $\Delta t$, but that they incur a breakdown in accuracy as $\Delta t$ increases. Further to this, we show in Section~\ref{sec:continvsdiscrete} that when the time-step is fixed to the value $\Delta t=1$, as in much of the literature, discrete-time approaches break down in accuracy when the transition rates ($\mu$ and $\beta$) are too large. This limits the range of parameters that can be studied with discrete-time approaches. We illustrate this with an example from the literature, also showing how synchronous updating simulation schemes can favour discrete-time formalisms leading to biased conclusions when comparing against continuous-time theories. Finally in Section~\ref{sec:effect_on_threshold} we show that overly-large values of $\tilde{\beta}$ and $\tilde{\mu}$ can affect the value of the epidemic threshold, even if the effective transition rate defined as $\gamma = \beta/\mu = \tilde{\beta}/\tilde{\mu}$ is small.

\section{Effect of the time-step $\Delta t$ when discretizing time.}
\label{sec:comparison}

\begin{figure*}
  \centering
  \begin{subfigure}{0.6\textwidth}
    \centering
    \includegraphics[width=\columnwidth]{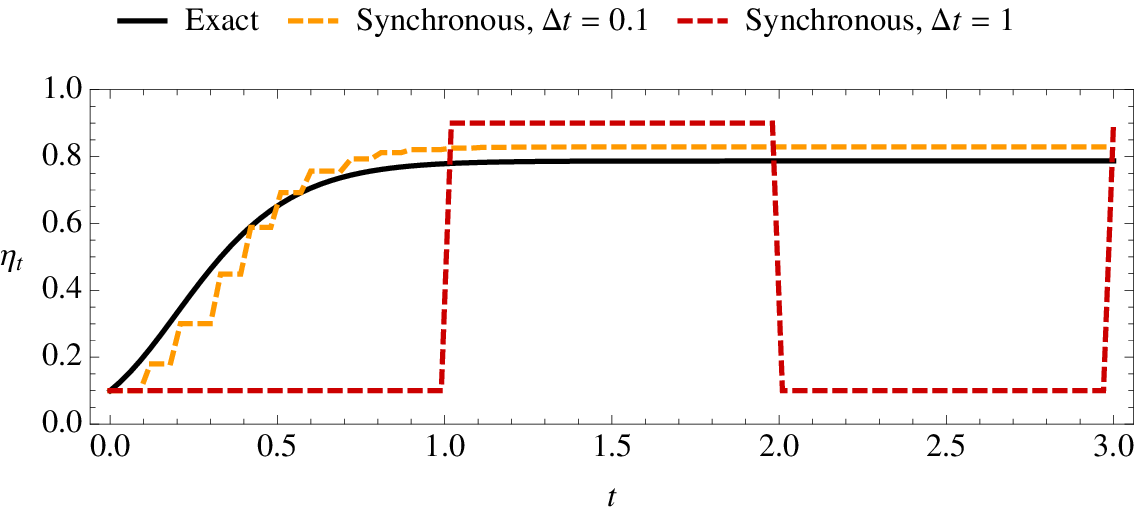}
    \caption{Time evolution of $\eta_t$}
    \label{fig:ex_evolution}
  \end{subfigure}
  \\  \vspace{10pt}

  \begin{subfigure}{0.3\textwidth}
    \includegraphics[width=\columnwidth]{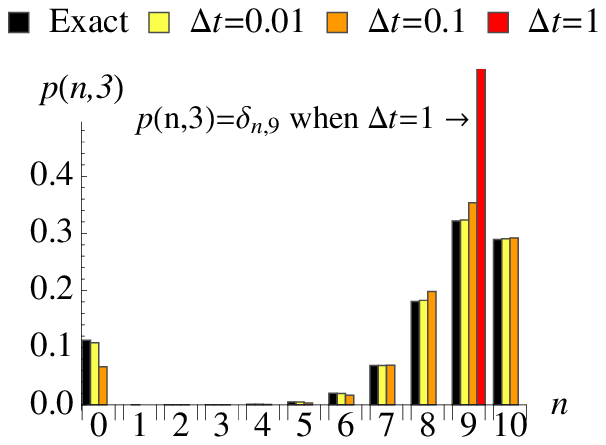}
    \caption{Synchronous updating}
    \label{fig:exact_S}
  \end{subfigure}
  \hspace{20pt}
  \begin{subfigure}{0.3\textwidth}
    \includegraphics[width=0.94\columnwidth]{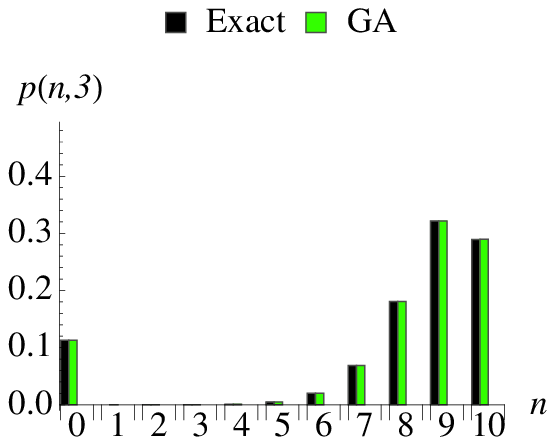}
    \caption{Gillespie algorithm}
    \label{fig:exact_GA}
  \end{subfigure}
  \caption{SIS dynamics with $\beta = \mu = 1$ on a complete graph with $N=10$ nodes. (a): Time evolution of the expected fraction of infected nodes $\eta_t$ for both the exact master equation Eq.~\eqref{eq:masterequation_complete} (solid line) and synchronous updating simulations with time steps $\Delta t=0.1$ and $\Delta t=1$ (dashed lines). The Gillespie algorithm and Exact curves are indistinguishable. (b) and (c): Histograms showing the exact probability mass
    function $p(z,t)$ at $t=3$ --- calculated from numerical integration of
    Eq.~(\ref{eq:masterequation_complete}) --- and the probability mass
    functions obtained empirically from $10^6$ simulation realizations for both synchronous updating simulations with time steps $\Delta t=0.01$, $\Delta t=0.1$ and $\Delta t=1$ and the Gillespie algorithm respectively.}
  \label{fig:comparison}
\end{figure*}

In this section, we analyze the discrete-time approximations introduced in Section~\ref{sec:discrete_SIS} as a function of the size of the discrete time-step $\Delta t$. We do this by carrying out synchronous updating simulations for various values of $\Delta t$ and comparing them against exact results obtained from the master equation. Numerical simulations are carried out in \verb!C++! and the code is available online from~\cite{website:SIScode}.


As our example, we consider SIS dynamics on a complete graph of $N$ nodes, i.e., a graph where every pair of nodes is connected. On such a graph, the SIS dynamics are defined by the rate functions
\begin{equation}
  r_i =
  \begin{cases}
    \;\beta Z_t & \mbox{if } X^i_t = S \\
    \;\mu & \mbox{if } X^i_t = I
  \end{cases},
  \label{eq:SIS}
\end{equation}
where $Z_t$ is the number of infected nodes at time $t$ and $\beta$ and $\mu$ are
the infection rate and recovery rate respectively, consistent with Eq.~\eqref{eq:transitionrateSIS} for the complete graph. We choose the complete graph because on such a graph the master equation given in Eq.~\eqref{eq:masterequation} can be reduced from a system of $2^N$ equations for $p(\textbf{y},t)$ to a system of $N+1$ equations for $p(n,t)$, the probability that there are $n$ infected nodes in the graph at time $t$~\cite{barrat2008dynamical}. This reduced system is given by
\begin{align}
  \frac{d}{dt}&p(n,t)=-\Big(\mu n + \beta n (N-n)\Big)p(n,t) \nonumber \\ &+ \mu(n+1)  p(n+1,t) + \beta(n-1)(N-n+1) p(n-1,t),
  \label{eq:masterequation_complete}
\end{align}
for $0\leq n \leq N$ with initial conditions $p(n,0) = p_{0}(n)$. For small values of $N$, this system can easily be solved using standard differential-equation solvers, giving us a gold standard against which to compare the discrete-time simulations. We also perform Gillespie algorithm simulations to illustrate the accuracy and the speed of such simulations and thus their efficacy in simulating continuous-time dynamics.

We present the results for SIS dynamics with $\beta=1$, $\mu=1$ running on a complete graph with $N=10$ nodes in Fig.~\ref{fig:comparison}. We plot the solution of Eq.~(\ref{eq:masterequation_complete}) as well as the numerical results given by the Gillespie algorithm and synchronous updating schemes with different time steps $\Delta t$. For the numerical simulations, we performed $10^6$ realisations and obtained the corresponding $p(n,t)$ by taking the fraction of realisations in which there are $n$ infected nodes at time $t$. For the synchronous updating simulations, we consider time steps of $\Delta t = 0.01, 0.1$ and $1$. From Fig.~\ref{fig:cdtapprox}, it is clear that these values of $\Delta t$ with $\mu=\beta=1$ will give a comprehensive range on which to judge the accuracy of the discrete-time approach.

\begin{table*}[t]
    \centering
    \begin{subtable}[t]{0.45 \columnwidth}
      \caption{Complete network (Section~\ref{sec:comparison})}
      \centering
      \begin{tabular}{c  c | c | c |}
              \hline
        \multicolumn{2}{|c }{Gillespie} &  \multicolumn{2}{| c |}{Synchronous}  \\
           \hline
        \hline
        \multicolumn{1}{|c }{\;\;\; $\Delta t$ \;\;\;} & \multicolumn{1}{|c |}{\;\;\; $T$\;\;\;} & \;\; $\Delta t$ \;\; & \;\; $T$ \;\;   \\
       \hline
       \multicolumn{1}{|c }{ - } & \multicolumn{1}{|c |}{5.67} &  0.01 & 65.62 \\
       \cline{1-2} &   & 0.1 & 12.46  \\
      &  & 1 & 4.48 \\
      \cline{3-4}
        \end{tabular}
        \label{tab:timing1}
    \end{subtable}
    \hspace{50pt}
    \begin{subtable}[t]{0.45 \columnwidth}
      \caption{\er network (Section~\ref{sec:continvsdiscrete})}
      \centering
      \begin{tabular}{c  c | c | c |}
        \hline
        \multicolumn{2}{|c }{Gillespie} &   \multicolumn{2}{| c |}{Synchronous}  \\
        \hline
        \hline
        \multicolumn{1}{|c }{\;\;\; $\Delta t$ \;\;\;} & \multicolumn{1}{|c |}{\;\;\; $T$\;\;\;} & \;\; $\Delta t$ \;\; & \;\; $T$ \;\;   \\
        \hline
        \multicolumn{1}{|c }{ - } & \multicolumn{1}{|c |}{42.1} &  0.01 & 2555.8 \\
        \cline{1-2} &  & 1 & 81.4  \\
        \cline{3-4}
      \end{tabular}
      \label{tab:timing2}
    \end{subtable}
    \caption{Time $T$ (in seconds) taken to carry out the Gillespie algorithm and synchronous updating numerical simulations for the examples described in Section~\ref{sec:comparison} (Table~\ref{tab:timing1}) and Section~\ref{sec:continvsdiscrete} (Table~\ref{tab:timing1}). The simulation code was written in C++ and the simulations were run on a single-CPU contemporary desktop computer.}
    \label{tab:timing}
\end{table*}

We consider the SIS process at time $t=3$ at which stage the expected fraction of infected nodes $\eta_t = \sum_{n=0}^{10}n p(n,t)$ has reached a metastable state (Fig~\ref{fig:ex_evolution})
~\footnote{We refer to the metastable state as the state when the expected fraction of infected nodes has reached a plateau which very slowly decreases towards 0.}.
At $t=3$ we empirically construct $p(n,3)$ from the synchronous updating simulations and compare it to $p(n,3)$ calculated from the master equation~\eqref{eq:masterequation_complete}. Figure~\ref{fig:exact_S} shows this comparison. From this Figure it is clear that while the discrete-time simulations are quite accurate for small $\Delta t$ this accuracy can fully break down when $\Delta t$ is too large. The accuracy of the probability distribution in the metastable state depends highly on the value of the time step used to reach the metastable state. In the synchronous updating simulations with $\Delta t=1$ the results are highly inaccurate with all of the probability concentrated on $n=9$, i.e., $p(n,3) = \delta_{n,9}$. Even the case $\Delta t=0.1$, while fairly accurate, shows discrepancies in both the probability distribution $p(n,3)$ and the expected fraction of infected nodes $\eta_t$ (Fig.~\ref{fig:ex_evolution}). Considering that the error $\epsilon$ between $\tilde{\mu}$ (resp. $\tilde{\beta}$) and $1-e^{-\mu \Delta t}$ ($1-e^{-\beta \Delta t}$) is less than 0.005 for $\mu=\beta=1$ and $\Delta t=0.1$ (Fig.~\ref{fig:cdtapprox}), we conclude that these discrepancies are due to the simultaneous state-changes in synchronous updating which are uncharacteristic of the continuous-time process.

In Fig.~\ref{fig:exact_GA} we compare $p(n,3)$ constructed empirically from the Gillespie algorithm to $p(n,3)$ calculated from the master equation. The Gillespie algorithm is extremely accurate and matches the exact $p(n,t)$ to a high degree of precision. Furthermore, this algorithm is computationally rapid. We performed a short comparison of the simulation algorithms in terms of speed, showing in Table~\ref{tab:timing} the simulation run times for the $10^6$ realisations for the Gillespie Algorithm and for synchronous
updating with various values of $\Delta t$. For $\Delta t=0.01$ -- corresponding to the simulations which most closely match the accuracy of the Gillespie simulations -- the Gillespie algorithm is an order of magnitude faster. This computational speed, along with its natural precision of the algorithm, make the Gillespie algorithm an optimal algorithm for simulating continuous-time dynamics.

To summarize, the accuracy of discrete-time approximations to continuous-time dynamics depends highly on the size of the discrete-time step $\Delta t$ at which the system evolves. This has extremely important implications for real-world predictive models of epidemic spreads that are discrete-time based~\cite{degli2008mitigation,balcan2010modeling}, as overly-large time-steps can affect the prediction of both the expected evolution of a contagion (Fig.~\ref{fig:ex_evolution}) as well as variance or confidence intervals around the expected evolution (Fig.~\ref{fig:exact_S}).

In the next section, we fix the time-step at $\Delta t=1$ and show how the accuracy breaks down when the infection and recovery rates are too large, showing that discrete-time formalisms using this approach are limited in the ranges of the rate parameters that they can study and thus their ability to match continuous-time dynamics.

\section{Limitations on range of parameter values when $\Delta t=1$}
\label{sec:continvsdiscrete}

As mentioned in Section~\ref{sec:discrete_SIS}, synchronous updating has the same characteristics of
discrete-time systems which are characterized by transition probabilities and difference equations of the form
\begin{equation}
  p(\textbf{y},t+\Delta t) = f\left(\{p(\textbf{y}',t)\}_{\textbf{y}'\in \Omega}\right),
  \label{eq:difference_eqs}
\end{equation}
where $p(\textbf{y},t+\Delta t)$ --- the probability that the system is in state $\textbf{y}$ at
time $t+\Delta t$ --- is a function of the probabilities $p(\textbf{y}',t)$ for all possible states $\textbf{y}'$ in the sample space $\Omega$. On the other hand, continuous-time systems are characterized by transition rates and differential equations of the form given by the master equation  Eq.~\eqref{eq:masterequation}. Although the discrete-time formulation coincides with the
continuous-time one in the limit $\Delta t\rightarrow 0$, the dynamics
will differ for non-infinitesimal $\Delta t$. Issues then arise when
comparing discrete and continuous-time systems and the choice of
numerical scheme becomes important. We illustrate this now with an example
from the literature, while also showing how the accuracy of discrete-time approaches with $\Delta t = 1$ can be insufficient for large values of the transition rates.

A prominent current strand
of research is the behaviour of the SIS model on infinite networks
with power-law degree
distributions~\cite{pastor2001epidemic,chakrabarti2008epidemic,
  castellano2010thresholds, goltsev2012localization, van2012epidemic, boguna2013nature, mata2015multiple}. In~\cite{chakrabarti2008epidemic},
Chakrabarti \emph{et al.} introduced the NonLinear Dynamical Systems (NLDS)
theory, a discrete-time approach to SIS modeling with a set of mean-field
difference equations of the form
\begin{equation}
  p_{i,t+1} = f\left(\{p_{j,t}\}_{j=1}^N\right),
  \label{eq:NDLS_eqn}
\end{equation}
for $0 \leq i \leq N$, where $p_{i,t+1}$ is the probability that node $i$ is infected at time $t$. They compare their results to two continuous-time formulations, the Heterogeneous Mean
Field (HMF) approach of Pastor-Sattoras and Vespignani~\cite{pastor2001epidemic} and the
Kephart-White (KW)~\cite{kephart1991directed} approach. The bases of the comparison are
synchronous updating numerical simulations with a time step $\Delta t=1$ and
it is found --- see for example Figure 4
of~\cite{chakrabarti2008epidemic} --- that the NLDS theory is much closer to the numerical
simulations than both the HMF and KW theories.

\begin{figure*}[!t]
    \centering
    \begin{subfigure}{0.35\textwidth}
      \caption{\;\;\;$\mu=0.48, \beta = 0.2$}
      \includegraphics[width = \columnwidth]{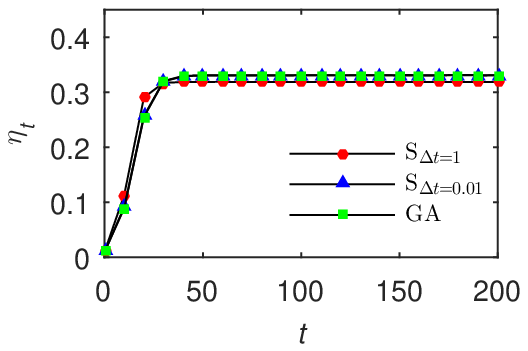}
      \label{fig:NLDS_mu48}
    \end{subfigure}
    \begin{subfigure}{0.35\textwidth}
      \caption{\;\;\;$\mu=0.72, \beta = 0.2$}
      \includegraphics[width = \columnwidth]{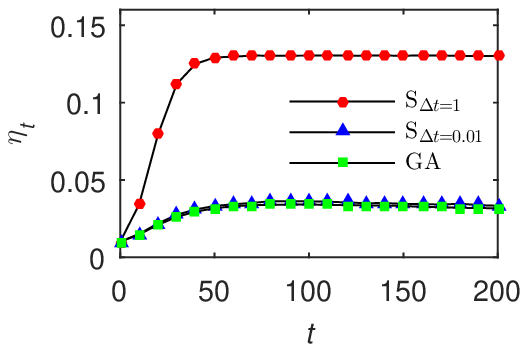}
      \label{fig:NLDS_mu72}
    \end{subfigure}
    \caption{Time evolution of the expected fraction of infected nodes $\eta_t$ for SIS dynamics on an \er network with 1000 nodes and average degree $\langle k \rangle = 4$. Each trajectory is averaged over $10^4$ realisations. Synchronous updating with a time step of $\Delta t=1$ --- as in~\cite{chakrabarti2008epidemic} --- is given by the circular symbols (S$_{\Delta t=1}$). This significantly deviates from Gillespie algorithm simulations (GA, triangular symbols) and synchronous updating simulations with a small time step of $\Delta t=0.01$ (S$_{\Delta t=0.01}$, square symbols) if the transition rates $\mu$ and/or $\beta$ are too large.}
  \label{fig:NLDS}
\end{figure*}

However, the comparison of discrete-time and continuous-time
formulations in this manner is biased. Synchronous updating with a time step $\Delta t=1$ is
the correct procedure for numerically simulating discrete-time
dynamics. On the other hand, to simulate continuous-time
dynamics either synchronous updating with a vanishingly small time
step or a continuous-time simulation scheme such as the Gillespie algorithm should be used.

To illustrate the difference resulting from the use of the
different updating methods we reproduce an example
from~\cite{chakrabarti2008epidemic}. The example is SIS dynamics on an
\er ~network of 1000 nodes and mean degree $\langle k \rangle=4$. Figure~\ref{fig:NLDS} shows various
numerical simulations of these dynamics. Again, the computer code used to
perform the simulations is available from~\cite{website:SIScode}. Included in Fig.~\ref{fig:NLDS} are synchronous
updating simulations with a time step $\Delta t=1$, as
in~\cite{chakrabarti2008epidemic}, along with synchronous updating
simulations with a small time step $\Delta t=0.01$ and Gillespie algorithm simulations. In Fig. 4(c) of~\cite{chakrabarti2008epidemic}, where $\mu=0.72$ and $\beta = 0.2$, it
can be seen that the fraction $\bar{\eta}$ of infected
nodes in the metastable state given by the NLDS theory matches very closely the synchronous updating numerical
simulations. However, as can be seen in Fig.~\ref{fig:NLDS_mu72} here, these
synchronous updating simulations differ quite significantly from continuous-time
simulations, which plateau at the metastable state with $\bar{\eta}\approx 0.04$. The KW theory,
which in~\cite{chakrabarti2008epidemic} is rejected as being inaccurate, actually converges to a value much closer to the continuous-time
simulations than the NLDS theory. Thus, using the correct simulation technique, the conclusions in~\cite{chakrabarti2008epidemic} should be reversed: the KW model is more accurate than the NDLS model.

For fixed $\Delta t=1$, the accuracy of the discrete-time approach decreases as $\mu$ and $\beta$ increase. In the example above, when $\mu$ is decreased from $\mu=0.72$ to $\mu=0.48$ the discrete-time simulations match relatively closer to the continuous-time simulations (Fig.~\ref{fig:NLDS_mu48}), while when $\mu$ is decreased further to $\mu=0.24$ the discrepancy between the two simulations in negligible. Chakrabarti \emph{et al.} state that their model ``outperforms (the KW model) when $\mu$ is high''. However the opposite is the case: their discrete-time approach breaks down in accuracy (as an approximation to the continuous-time process) as $\mu$ increases.

We conclude with an observation to motivate the next section. As $\mu$ is increased from $\mu=0.48$ (Fig.~\ref{fig:NLDS_mu48}) to $\mu=0.72$ (Fig.~\ref{fig:NLDS_mu72}), the fraction of infected nodes in the metastable state $\bar{\eta}$ (at, for example, $t=100$) decreases for both the continuous-time simulations and discrete-time simulations. However, $\bar{\eta}$ decreases quicker for the continuous-time simulations and so it would seem that the critical value $\mu_c$ at which $\bar{\eta}$ first becomes zero will be different depending on whether a discrete-time or continuous-time approach is used. This has implications for the epidemic threshold, which is the focus of the next section.

\section{Effect on Epidemic threshold.}
\label{sec:effect_on_threshold}

A characteristic of SIS dynamics is the occurrence of phase transitions as the effective transition rate $\gamma$ is varied. Recall that the effective transition rate is defined as the ratio of the infection rate to the recovery rate, i.e., $\gamma = \beta/\mu$. Depending on the structure of the network and whether the network is finite or infinite, the critical point, or epidemic threshold, $\gamma_c$ between different phases can vary. As mentioned in Section~\ref{sec:continvsdiscrete}, there are still remaining questions about the steady-state behaviour of the SIS model --- particularly the value of the epidemic threshold on such networks --- and so a good understanding of how different approximations affect the value of the epidemic threshold is important. In this section, we show that although the epidemic threshold is defined in terms of the ratio $\gamma = \beta/\mu = \tilde{\beta}/\tilde{\mu}$, the individual values of the transition probabilities $\tilde{\beta}$ and $\tilde{\mu}$ used in discrete-time approaches affect the value of the epidemic threshold when it is calculated by (a) performing discrete-time numerical simulations or (b) iterating a discrete-time system (such as Eq.~\eqref{eq:difference_eqs}) from a set of initial conditions (as, for example, in~\cite{gomez2010discrete}). Note however that the epidemic threshold predicted by steady-state analysis (i.e., setting $p_{t+1} = p_{t}$ in Eq.~\eqref{eq:NDLS_eqn}) --- such as in \cite{chakrabarti2008epidemic} --- is completely valid.

We show how $\tilde{\mu}$ and $\tilde{\beta}$ affect the value of the epidemic threshold in the following manner. For a given network, we fix the value of $\tilde{\mu}$ and vary $\tilde{\beta}$ so that the effective transition rate $\gamma$ varies between $\gamma_{min}$ and $\gamma_{max}$ where $\gamma_{min}$ and $\gamma_{max}$ are chosen so that the epidemic threshold lies between them, i.e.,  $\gamma_{min} \ll \gamma_c \ll \gamma_{max}$. Thus when $\tilde{\mu}$ is small (resp. large), $\tilde{\beta}$ will be small (large) so that their ratio lies in the range $\gamma_{min} \ll \tilde{\beta}/\tilde{\mu} \ll \gamma_{max}$. We perform standard synchronous updating simulations (with $\Delta t=1$) and obtain the critical value $\gamma_c$ as the smallest value of $\gamma$ such that the fraction of nodes in the metastable state is non-zero. If the epidemic threshold depends only on the ratio $\gamma=\tilde{\beta}/\tilde{\mu}$ and is independent of the individual values of $\tilde{\mu}$ and $\tilde{\beta}$, then $\gamma_c$ should be the same regardless of the value of $\tilde{\mu}$ which is fixed. However, we find that this is not the case.

\begin{figure}[!t]
  \centering
  \includegraphics[width = 0.75\columnwidth]{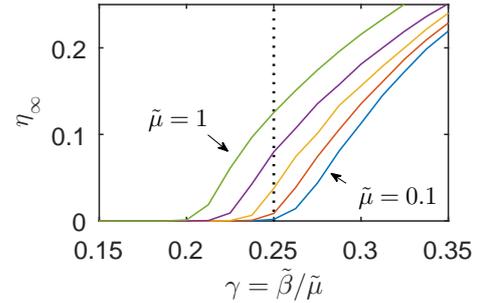}
  \caption{The fraction of infected nodes $\eta_{\infty}$ in the metastable state in an \er network with $10^4$ nodes and mean- degree $\langle k \rangle = 4$ for various values of the transition probabilities $\tilde{\mu}$ and $\tilde{\beta}$. Along each of the five curves shown, $\tilde{\mu}$ is fixed at the values 0.1, 0.325, 0.55, 0.775 and 1 respectively and $\tilde{\beta}$ varies so that $\gamma = \tilde{\beta}/\tilde{\mu}$ varies between 0.15 and 0.35. The vertical dashed line indicates the epidemic threshold $\gamma_c = 0.25$ predicted by both the NLDS and HMF theories.}
  \label{fig:ET_NLDS}
\end{figure}

We perform this experiment on an \er network with $N=10^4$ nodes and mean degree $\langle k \rangle = 4$, the same network as used in the example of Section~\ref{sec:continvsdiscrete} (Fig.~\ref{fig:ET_NLDS}). On such a network, the epidemic threshold is predicted by steady-state analysis of both the NDLS and HMF theories as $\gamma_c = 0.25$. From Fig.~\ref{fig:ET_NLDS} we see that when $\tilde{\mu}$ is small ($\tilde{\mu}$=0.1), the epidemic threshold predicted by synchronous updating simulations corresponds to this value $\gamma_c = 0.25$. However as $\tilde{\mu}$ (and thus $\tilde{\beta}$) increases, the accuracy of the discrete-time approach breaks down and both the fraction of infected nodes in the metastable state and the epidemic threshold deviate from the true values. The epidemic threshold decreases from $\gamma_c = 0.25$ when $\tilde{\mu} = 0.1$ to $\gamma_c = 0.2$ when $\tilde{\mu} = 1$, even though the ratio $\gamma = \tilde{\beta}/\tilde{\mu}$ remains in the same range. Thus in discrete-time formalisms the steady-state behaviour is not fully determined by the effective transition rate $\gamma$ but also depends on $\tilde{\mu}$ and $\tilde{\beta}$. From our analysis in Section~\ref{sec:comparison} (Fig.~\ref{fig:ex_evolution}) it is clear that the metastable state reached iteratively from an initial condition depends on the single-step transition probabilities $\tilde{\mu}$ and $\tilde{\beta}$. If these are too large, the errors introduced in the discrete-time approximation become significant, affecting the metastable state and the value of the epidemic threshold.

The results of this Section have important implications for discrete-time approaches. Firstly, they show that the epidemic threshold calculated empirically using synchronous updating simulations can be incorrect if $\tilde{\mu}$ and $\tilde{\beta}$ are too large, even if the ratio between them is small. Secondly, they have implications for calculating the epidemic threshold from discrete-time systems of the form $p_{t+1} = f(p_{t})$ by forward iterating the system from an initial condition~\cite{gomez2010discrete}. If the transition probabilities $\tilde{\mu}$ and $\tilde{\beta}$ used in such systems are too large then the metastable state will be affected, possibly leading to a miscalculation of the epidemic threshold.

\section{Conclusions}

In this paper, we have provided conclusive evidence of the limitations of discrete-time approaches as approximations to continuous-time contagion processes. When the state transition probabilities are too large, such approaches become inaccurate and misrepresentative of the underlying continuous-time processes, thus compromising their utility and their applicability to prediction and analysis.

Our message is clear - due care needs to be taken when implementing discrete-time methods as approximations to continuous-time dynamical processes. Being constructive, we have briefly discussed alternatives. For simulations of continuous-time processes on networks, event-based simulations such as the Gillespie algorithm are more favourable than synchronous updating schemes both in terms of accuracy and speed. For theoretical analysis, continuous-time analogues~\cite{van2009virus,goltsev2012localization} of discrete-time approaches should be employed because they are unconstrained in the range of dynamics parameter values which can be studied.

\section*{Acknowledgment}

This work has been partially funded by Science Foundation Ireland, Grant Nos. 11/PI/1026 and 12/PI/1683, and the FET-Proactive project PLEXMATH (FP7-ICT-2011-8; Grant No. 317614), funded by the European Commission.

\bibliographystyle{apsrev4-1}
\bibliography{library}

\begin{thebibliography}{63}%
\makeatletter
\providecommand \@ifxundefined [1]{%
 \@ifx{#1\undefined}
}%
\providecommand \@ifnum [1]{%
 \ifnum #1\expandafter \@firstoftwo
 \else \expandafter \@secondoftwo
 \fi
}%
\providecommand \@ifx [1]{%
 \ifx #1\expandafter \@firstoftwo
 \else \expandafter \@secondoftwo
 \fi
}%
\providecommand \natexlab [1]{#1}%
\providecommand \enquote  [1]{``#1''}%
\providecommand \bibnamefont  [1]{#1}%
\providecommand \bibfnamefont [1]{#1}%
\providecommand \citenamefont [1]{#1}%
\providecommand \href@noop [0]{\@secondoftwo}%
\providecommand \href [0]{\begingroup \@sanitize@url \@href}%
\providecommand \@href[1]{\@@startlink{#1}\@@href}%
\providecommand \@@href[1]{\endgroup#1\@@endlink}%
\providecommand \@sanitize@url [0]{\catcode `\\12\catcode `\$12\catcode
  `\&12\catcode `\#12\catcode `\^12\catcode `\_12\catcode `\%12\relax}%
\providecommand \@@startlink[1]{}%
\providecommand \@@endlink[0]{}%
\providecommand \url  [0]{\begingroup\@sanitize@url \@url }%
\providecommand \@url [1]{\endgroup\@href {#1}{\urlprefix }}%
\providecommand \urlprefix  [0]{URL }%
\providecommand \Eprint [0]{\href }%
\providecommand \doibase [0]{http://dx.doi.org/}%
\providecommand \selectlanguage [0]{\@gobble}%
\providecommand \bibinfo  [0]{\@secondoftwo}%
\providecommand \bibfield  [0]{\@secondoftwo}%
\providecommand \translation [1]{[#1]}%
\providecommand \BibitemOpen [0]{}%
\providecommand \bibitemStop [0]{}%
\providecommand \bibitemNoStop [0]{.\EOS\space}%
\providecommand \EOS [0]{\spacefactor3000\relax}%
\providecommand \BibitemShut  [1]{\csname bibitem#1\endcsname}%
\let\auto@bib@innerbib\@empty
\bibitem [{\citenamefont {Newman}(2010)}]{newman2010networks}%
  \BibitemOpen
  \bibfield  {author} {\bibinfo {author} {\bibfnamefont {M.~E.}\ \bibnamefont
  {Newman}},\ }\href@noop {} {\emph {\bibinfo {title} {Networks: {A}n
  {I}ntroduction}}}\ (\bibinfo  {publisher} {Oxford University Press},\
  \bibinfo {year} {2010})\BibitemShut {NoStop}%
\bibitem [{\citenamefont {Pastor-Satorras}\ and\ \citenamefont
  {Vespignani}(2001)}]{pastor2001epidemic}%
  \BibitemOpen
  \bibfield  {author} {\bibinfo {author} {\bibfnamefont {R.}~\bibnamefont
  {Pastor-Satorras}}\ and\ \bibinfo {author} {\bibfnamefont {A.}~\bibnamefont
  {Vespignani}},\ }\href@noop {} {\bibfield  {journal} {\bibinfo  {journal}
  {Physical Review Letters}\ }\textbf {\bibinfo {volume} {86}},\ \bibinfo
  {pages} {3200} (\bibinfo {year} {2001})}\BibitemShut {NoStop}%
\bibitem [{\citenamefont {Anderson}\ and\ \citenamefont
  {May}(1979)}]{anderson1979population}%
  \BibitemOpen
  \bibfield  {author} {\bibinfo {author} {\bibfnamefont {R.~M.}\ \bibnamefont
  {Anderson}}\ and\ \bibinfo {author} {\bibfnamefont {R.~M.}\ \bibnamefont
  {May}},\ }\href@noop {} {\bibfield  {journal} {\bibinfo  {journal} {Nature}\
  ,\ \bibinfo {pages} {361}} (\bibinfo {year} {1979})}\BibitemShut {NoStop}%
\bibitem [{\citenamefont {Axelrod}(1997)}]{axelrod1997dissemination}%
  \BibitemOpen
  \bibfield  {author} {\bibinfo {author} {\bibfnamefont {R.}~\bibnamefont
  {Axelrod}},\ }\href@noop {} {\bibfield  {journal} {\bibinfo  {journal}
  {Journal of {C}onflict {R}esolution}\ }\textbf {\bibinfo {volume} {41}},\
  \bibinfo {pages} {203} (\bibinfo {year} {1997})}\BibitemShut {NoStop}%
\bibitem [{\citenamefont {Watts}(2002)}]{watts2002simple}%
  \BibitemOpen
  \bibfield  {author} {\bibinfo {author} {\bibfnamefont {D.~J.}\ \bibnamefont
  {Watts}},\ }\href@noop {} {\bibfield  {journal} {\bibinfo  {journal}
  {Proceedings of the {N}ational {A}cademy of {S}ciences}\ }\textbf {\bibinfo
  {volume} {99}},\ \bibinfo {pages} {5766} (\bibinfo {year}
  {2002})}\BibitemShut {NoStop}%
\bibitem [{\citenamefont {O'Sullivan}\ \emph {et~al.}(2015)\citenamefont
  {O'Sullivan}, \citenamefont {O'Keeffe}, \citenamefont {Fennell},\ and\
  \citenamefont {Gleeson}}]{o2015mathematical}%
  \BibitemOpen
  \bibfield  {author} {\bibinfo {author} {\bibfnamefont {D.~J.}\ \bibnamefont
  {O'Sullivan}}, \bibinfo {author} {\bibfnamefont {G.~J.}\ \bibnamefont
  {O'Keeffe}}, \bibinfo {author} {\bibfnamefont {P.~G.}\ \bibnamefont
  {Fennell}}, \ and\ \bibinfo {author} {\bibfnamefont {J.~P.}\ \bibnamefont
  {Gleeson}},\ }\href@noop {} {\bibfield  {journal} {\bibinfo  {journal}
  {Frontiers in Physics}\ }\textbf {\bibinfo {volume} {3}},\ \bibinfo {pages}
  {71} (\bibinfo {year} {2015})}\BibitemShut {NoStop}%
\bibitem [{\citenamefont {Centola}(2010)}]{centola2010spread}%
  \BibitemOpen
  \bibfield  {author} {\bibinfo {author} {\bibfnamefont {D.}~\bibnamefont
  {Centola}},\ }\href@noop {} {\bibfield  {journal} {\bibinfo  {journal}
  {Science}\ }\textbf {\bibinfo {volume} {329}},\ \bibinfo {pages} {1194}
  (\bibinfo {year} {2010})}\BibitemShut {NoStop}%
\bibitem [{\citenamefont {Hodas}\ and\ \citenamefont
  {Lerman}(2014)}]{hodas2014simple}%
  \BibitemOpen
  \bibfield  {author} {\bibinfo {author} {\bibfnamefont {N.~O.}\ \bibnamefont
  {Hodas}}\ and\ \bibinfo {author} {\bibfnamefont {K.}~\bibnamefont {Lerman}},\
  }\href@noop {} {\bibfield  {journal} {\bibinfo  {journal} {Scientific
  {R}eports}\ }\textbf {\bibinfo {volume} {4}} (\bibinfo {year}
  {2014})}\BibitemShut {NoStop}%
\bibitem [{\citenamefont {Gleeson}\ \emph
  {et~al.}(2012{\natexlab{a}})\citenamefont {Gleeson}, \citenamefont {Hurd},
  \citenamefont {Melnik},\ and\ \citenamefont {Hackett}}]{gleeson2012systemic}%
  \BibitemOpen
  \bibfield  {author} {\bibinfo {author} {\bibfnamefont {J.~P.}\ \bibnamefont
  {Gleeson}}, \bibinfo {author} {\bibfnamefont {T.}~\bibnamefont {Hurd}},
  \bibinfo {author} {\bibfnamefont {S.}~\bibnamefont {Melnik}}, \ and\ \bibinfo
  {author} {\bibfnamefont {A.}~\bibnamefont {Hackett}},\ }in\ \href@noop {}
  {\emph {\bibinfo {booktitle} {Advances in Network Analysis and its
  Applications}}}\ (\bibinfo  {publisher} {Springer},\ \bibinfo {year} {2012})\
  pp.\ \bibinfo {pages} {27--56}\BibitemShut {NoStop}%
\bibitem [{\citenamefont {Gai}\ and\ \citenamefont
  {Kapadia}(2010)}]{gai2010contagion}%
  \BibitemOpen
  \bibfield  {author} {\bibinfo {author} {\bibfnamefont {P.}~\bibnamefont
  {Gai}}\ and\ \bibinfo {author} {\bibfnamefont {S.}~\bibnamefont {Kapadia}},\
  }in\ \href@noop {} {\emph {\bibinfo {booktitle} {Proceedings of the Royal
  Society of London A: Mathematical, Physical and Engineering Sciences}}}\
  (\bibinfo {organization} {The Royal Society},\ \bibinfo {year} {2010})\ p.\
  \bibinfo {pages} {rspa20090410}\BibitemShut {NoStop}%
\bibitem [{\citenamefont {May}\ and\ \citenamefont
  {Arinaminpathy}(2010)}]{may2010systemic}%
  \BibitemOpen
  \bibfield  {author} {\bibinfo {author} {\bibfnamefont {R.~M.}\ \bibnamefont
  {May}}\ and\ \bibinfo {author} {\bibfnamefont {N.}~\bibnamefont
  {Arinaminpathy}},\ }\href@noop {} {\bibfield  {journal} {\bibinfo  {journal}
  {Journal of the Royal Society Interface}\ }\textbf {\bibinfo {volume} {7}},\
  \bibinfo {pages} {823} (\bibinfo {year} {2010})}\BibitemShut {NoStop}%
\bibitem [{\citenamefont {Gleeson}\ \emph
  {et~al.}(2014{\natexlab{a}})\citenamefont {Gleeson}, \citenamefont {Cellai},
  \citenamefont {Onnela}, \citenamefont {Porter},\ and\ \citenamefont
  {Reed-Tsochas}}]{gleeson2014simple}%
  \BibitemOpen
  \bibfield  {author} {\bibinfo {author} {\bibfnamefont {J.~P.}\ \bibnamefont
  {Gleeson}}, \bibinfo {author} {\bibfnamefont {D.}~\bibnamefont {Cellai}},
  \bibinfo {author} {\bibfnamefont {J.-P.}\ \bibnamefont {Onnela}}, \bibinfo
  {author} {\bibfnamefont {M.~A.}\ \bibnamefont {Porter}}, \ and\ \bibinfo
  {author} {\bibfnamefont {F.}~\bibnamefont {Reed-Tsochas}},\ }\href@noop {}
  {\bibfield  {journal} {\bibinfo  {journal} {Proceedings of the National
  Academy of Sciences}\ }\textbf {\bibinfo {volume} {111}},\ \bibinfo {pages}
  {10411} (\bibinfo {year} {2014}{\natexlab{a}})}\BibitemShut {NoStop}%
\bibitem [{\citenamefont {Melnik}\ \emph {et~al.}(2013)\citenamefont {Melnik},
  \citenamefont {Ward}, \citenamefont {Gleeson},\ and\ \citenamefont
  {Porter}}]{melnik2013}%
  \BibitemOpen
  \bibfield  {author} {\bibinfo {author} {\bibfnamefont {S.}~\bibnamefont
  {Melnik}}, \bibinfo {author} {\bibfnamefont {J.~A.}\ \bibnamefont {Ward}},
  \bibinfo {author} {\bibfnamefont {J.~P.}\ \bibnamefont {Gleeson}}, \ and\
  \bibinfo {author} {\bibfnamefont {M.~A.}\ \bibnamefont {Porter}},\
  }\href@noop {} {\bibfield  {journal} {\bibinfo  {journal} {Chaos}\ }\textbf
  {\bibinfo {volume} {23}},\ \bibinfo {pages} {013124} (\bibinfo {year}
  {2013})}\BibitemShut {NoStop}%
\bibitem [{\citenamefont {Gleeson}\ \emph
  {et~al.}(2014{\natexlab{b}})\citenamefont {Gleeson}, \citenamefont {Ward},
  \citenamefont {O�{S}ullivan},\ and\ \citenamefont
  {Lee}}]{gleeson2014competition}%
  \BibitemOpen
  \bibfield  {author} {\bibinfo {author} {\bibfnamefont {J.~P.}\ \bibnamefont
  {Gleeson}}, \bibinfo {author} {\bibfnamefont {J.~A.}\ \bibnamefont {Ward}},
  \bibinfo {author} {\bibfnamefont {K.~P.}\ \bibnamefont {O�{S}ullivan}}, \
  and\ \bibinfo {author} {\bibfnamefont {W.~T.}\ \bibnamefont {Lee}},\
  }\href@noop {} {\bibfield  {journal} {\bibinfo  {journal} {Physical {R}eview
  {L}etters}\ }\textbf {\bibinfo {volume} {112}},\ \bibinfo {pages} {048701}
  (\bibinfo {year} {2014}{\natexlab{b}})}\BibitemShut {NoStop}%
\bibitem [{\citenamefont {Gleeson}(2011)}]{gleeson2011high}%
  \BibitemOpen
  \bibfield  {author} {\bibinfo {author} {\bibfnamefont {J.~P.}\ \bibnamefont
  {Gleeson}},\ }\href@noop {} {\bibfield  {journal} {\bibinfo  {journal}
  {Physical {R}eview {L}etters}\ }\textbf {\bibinfo {volume} {107}},\ \bibinfo
  {pages} {068701} (\bibinfo {year} {2011})}\BibitemShut {NoStop}%
\bibitem [{\citenamefont {Gleeson}(2013)}]{gleeson2013}%
  \BibitemOpen
  \bibfield  {author} {\bibinfo {author} {\bibfnamefont {J.~P.}\ \bibnamefont
  {Gleeson}},\ }\href {\doibase 10.1103/PhysRevX.3.021004} {\bibfield
  {journal} {\bibinfo  {journal} {Physical Review X}\ }\textbf {\bibinfo
  {volume} {3}},\ \bibinfo {pages} {021004} (\bibinfo {year} {2013})},\ \Eprint
  {http://arxiv.org/abs/1209.2983} {1209.2983} \BibitemShut {NoStop}%
\bibitem [{\citenamefont {Gleeson}\ \emph
  {et~al.}(2012{\natexlab{b}})\citenamefont {Gleeson}, \citenamefont {Melnik},
  \citenamefont {Ward}, \citenamefont {Porter},\ and\ \citenamefont
  {Mucha}}]{gleeson2012accuracy}%
  \BibitemOpen
  \bibfield  {author} {\bibinfo {author} {\bibfnamefont {J.~P.}\ \bibnamefont
  {Gleeson}}, \bibinfo {author} {\bibfnamefont {S.}~\bibnamefont {Melnik}},
  \bibinfo {author} {\bibfnamefont {J.~A.}\ \bibnamefont {Ward}}, \bibinfo
  {author} {\bibfnamefont {M.~A.}\ \bibnamefont {Porter}}, \ and\ \bibinfo
  {author} {\bibfnamefont {P.~J.}\ \bibnamefont {Mucha}},\ }\href@noop {}
  {\bibfield  {journal} {\bibinfo  {journal} {Physical {R}eview {E}}\ }\textbf
  {\bibinfo {volume} {85}},\ \bibinfo {pages} {026106} (\bibinfo {year}
  {2012}{\natexlab{b}})}\BibitemShut {NoStop}%
\bibitem [{\citenamefont {Melnik}\ \emph {et~al.}(2011)\citenamefont {Melnik},
  \citenamefont {Hackett}, \citenamefont {Porter}, \citenamefont {Mucha},\ and\
  \citenamefont {Gleeson}}]{melnik2011unreasonable}%
  \BibitemOpen
  \bibfield  {author} {\bibinfo {author} {\bibfnamefont {S.}~\bibnamefont
  {Melnik}}, \bibinfo {author} {\bibfnamefont {A.}~\bibnamefont {Hackett}},
  \bibinfo {author} {\bibfnamefont {M.~A.}\ \bibnamefont {Porter}}, \bibinfo
  {author} {\bibfnamefont {P.~J.}\ \bibnamefont {Mucha}}, \ and\ \bibinfo
  {author} {\bibfnamefont {J.~P.}\ \bibnamefont {Gleeson}},\ }\href@noop {}
  {\bibfield  {journal} {\bibinfo  {journal} {Physical Review E}\ }\textbf
  {\bibinfo {volume} {83}},\ \bibinfo {pages} {036112} (\bibinfo {year}
  {2011})}\BibitemShut {NoStop}%
\bibitem [{\citenamefont {Faqeeh}\ \emph {et~al.}(2015)\citenamefont {Faqeeh},
  \citenamefont {Melnik},\ and\ \citenamefont {Gleeson}}]{faqeeh2015network}%
  \BibitemOpen
  \bibfield  {author} {\bibinfo {author} {\bibfnamefont {A.}~\bibnamefont
  {Faqeeh}}, \bibinfo {author} {\bibfnamefont {S.}~\bibnamefont {Melnik}}, \
  and\ \bibinfo {author} {\bibfnamefont {J.~P.}\ \bibnamefont {Gleeson}},\
  }\href@noop {} {\bibfield  {journal} {\bibinfo  {journal} {Physical Review
  E}\ }\textbf {\bibinfo {volume} {91}},\ \bibinfo {pages} {052807} (\bibinfo
  {year} {2015})}\BibitemShut {NoStop}%
\bibitem [{\citenamefont {Gomes}\ \emph {et~al.}(2014)\citenamefont {Gomes},
  \citenamefont {Piontti}, \citenamefont {Rossi}, \citenamefont {Chao},
  \citenamefont {Longini}, \citenamefont {Halloran},\ and\ \citenamefont
  {Vespignani}}]{gomes2014assessing}%
  \BibitemOpen
  \bibfield  {author} {\bibinfo {author} {\bibfnamefont {M.~F.}\ \bibnamefont
  {Gomes}}, \bibinfo {author} {\bibfnamefont {A.}~\bibnamefont {Piontti}},
  \bibinfo {author} {\bibfnamefont {L.}~\bibnamefont {Rossi}}, \bibinfo
  {author} {\bibfnamefont {D.}~\bibnamefont {Chao}}, \bibinfo {author}
  {\bibfnamefont {I.}~\bibnamefont {Longini}}, \bibinfo {author} {\bibfnamefont
  {M.~E.}\ \bibnamefont {Halloran}}, \ and\ \bibinfo {author} {\bibfnamefont
  {A.}~\bibnamefont {Vespignani}},\ }\href@noop {} {\bibfield  {journal}
  {\bibinfo  {journal} {PLOS Currents Outbreaks}\ }\textbf {\bibinfo {volume}
  {1}} (\bibinfo {year} {2014})}\BibitemShut {NoStop}%
\bibitem [{\citenamefont {Gonz{\'a}lez-Bail{\'o}n}\ \emph
  {et~al.}(2011)\citenamefont {Gonz{\'a}lez-Bail{\'o}n}, \citenamefont
  {Borge-Holthoefer}, \citenamefont {Rivero},\ and\ \citenamefont
  {Moreno}}]{gonzalez2011dynamics}%
  \BibitemOpen
  \bibfield  {author} {\bibinfo {author} {\bibfnamefont {S.}~\bibnamefont
  {Gonz{\'a}lez-Bail{\'o}n}}, \bibinfo {author} {\bibfnamefont
  {J.}~\bibnamefont {Borge-Holthoefer}}, \bibinfo {author} {\bibfnamefont
  {A.}~\bibnamefont {Rivero}}, \ and\ \bibinfo {author} {\bibfnamefont
  {Y.}~\bibnamefont {Moreno}},\ }\href@noop {} {\bibfield  {journal} {\bibinfo
  {journal} {Scientific {R}eports}\ }\textbf {\bibinfo {volume} {1}} (\bibinfo
  {year} {2011})}\BibitemShut {NoStop}%
\bibitem [{\citenamefont {Lazer}\ \emph {et~al.}(2009)\citenamefont {Lazer},
  \citenamefont {Pentland}, \citenamefont {Adamic}, \citenamefont {Aral},
  \citenamefont {Barab{\'a}si}, \citenamefont {Brewer}, \citenamefont
  {Christakis}, \citenamefont {Contractor}, \citenamefont {Fowler},
  \citenamefont {Gutmann} \emph {et~al.}}]{lazer2009life}%
  \BibitemOpen
  \bibfield  {author} {\bibinfo {author} {\bibfnamefont {D.}~\bibnamefont
  {Lazer}}, \bibinfo {author} {\bibfnamefont {A.~S.}\ \bibnamefont {Pentland}},
  \bibinfo {author} {\bibfnamefont {L.}~\bibnamefont {Adamic}}, \bibinfo
  {author} {\bibfnamefont {S.}~\bibnamefont {Aral}}, \bibinfo {author}
  {\bibfnamefont {A.~L.}\ \bibnamefont {Barab{\'a}si}}, \bibinfo {author}
  {\bibfnamefont {D.}~\bibnamefont {Brewer}}, \bibinfo {author} {\bibfnamefont
  {N.}~\bibnamefont {Christakis}}, \bibinfo {author} {\bibfnamefont
  {N.}~\bibnamefont {Contractor}}, \bibinfo {author} {\bibfnamefont
  {J.}~\bibnamefont {Fowler}}, \bibinfo {author} {\bibfnamefont
  {M.}~\bibnamefont {Gutmann}},  \emph {et~al.},\ }\href@noop {} {\bibfield
  {journal} {\bibinfo  {journal} {Science (New York, NY)}\ }\textbf {\bibinfo
  {volume} {323}},\ \bibinfo {pages} {721} (\bibinfo {year}
  {2009})}\BibitemShut {NoStop}%
\bibitem [{\citenamefont {Van~Mieghem}(2014)}]{van2014performance}%
  \BibitemOpen
  \bibfield  {author} {\bibinfo {author} {\bibfnamefont {P.}~\bibnamefont
  {Van~Mieghem}},\ }\href@noop {} {\emph {\bibinfo {title} {Performance
  {A}nalysis of {C}omplex {N}etworks and {S}ystems}}}\ (\bibinfo  {publisher}
  {Cambridge University Press},\ \bibinfo {year} {2014})\BibitemShut {NoStop}%
\bibitem [{\citenamefont {Barrat}\ \emph {et~al.}(2008)\citenamefont {Barrat},
  \citenamefont {Barthelemy},\ and\ \citenamefont
  {Vespignani}}]{barrat2008dynamical}%
  \BibitemOpen
  \bibfield  {author} {\bibinfo {author} {\bibfnamefont {A.}~\bibnamefont
  {Barrat}}, \bibinfo {author} {\bibfnamefont {M.}~\bibnamefont {Barthelemy}},
  \ and\ \bibinfo {author} {\bibfnamefont {A.}~\bibnamefont {Vespignani}},\
  }\href@noop {} {\emph {\bibinfo {title} {{D}ynamical {P}rocesses on {C}omplex
  {N}etworks}}}\ (\bibinfo  {publisher} {Cambridge University Press
  Cambridge},\ \bibinfo {year} {2008})\BibitemShut {NoStop}%
\bibitem [{\citenamefont {Cox}\ and\ \citenamefont
  {Miller}(1977)}]{cox1977theory}%
  \BibitemOpen
  \bibfield  {author} {\bibinfo {author} {\bibfnamefont {D.~R.}\ \bibnamefont
  {Cox}}\ and\ \bibinfo {author} {\bibfnamefont {H.~D.}\ \bibnamefont
  {Miller}},\ }\href@noop {} {\emph {\bibinfo {title} {The {T}heory of
  {S}tochastic {P}rocesses}}}\ (\bibinfo  {publisher} {CRC Press},\ \bibinfo
  {year} {1977})\BibitemShut {NoStop}%
\bibitem [{\citenamefont {Kermack}\ and\ \citenamefont
  {McKendrick}(1927)}]{kermack1927contribution}%
  \BibitemOpen
  \bibfield  {author} {\bibinfo {author} {\bibfnamefont {W.~O.}\ \bibnamefont
  {Kermack}}\ and\ \bibinfo {author} {\bibfnamefont {A.~G.}\ \bibnamefont
  {McKendrick}},\ }in\ \href@noop {} {\emph {\bibinfo {booktitle} {Proceedings
  of the Royal Society of London A: Mathematical, Physical and Engineering
  Sciences}}},\ Vol.\ \bibinfo {volume} {115}\ (\bibinfo {organization} {The
  Royal Society},\ \bibinfo {year} {1927})\ pp.\ \bibinfo {pages}
  {700--721}\BibitemShut {NoStop}%
\bibitem [{\citenamefont {Toole}\ \emph {et~al.}(2012)\citenamefont {Toole},
  \citenamefont {Cha},\ and\ \citenamefont {Gonz{\'a}lez}}]{toole2012modeling}%
  \BibitemOpen
  \bibfield  {author} {\bibinfo {author} {\bibfnamefont {J.~L.}\ \bibnamefont
  {Toole}}, \bibinfo {author} {\bibfnamefont {M.}~\bibnamefont {Cha}}, \ and\
  \bibinfo {author} {\bibfnamefont {M.~C.}\ \bibnamefont {Gonz{\'a}lez}},\
  }\href@noop {} {\bibfield  {journal} {\bibinfo  {journal} {PloS {O}ne}\
  }\textbf {\bibinfo {volume} {7}},\ \bibinfo {pages} {e29528} (\bibinfo {year}
  {2012})}\BibitemShut {NoStop}%
\bibitem [{\citenamefont {Kitsak}\ \emph {et~al.}(2010)\citenamefont {Kitsak},
  \citenamefont {Gallos}, \citenamefont {Havlin}, \citenamefont {Liljeros},
  \citenamefont {Muchnik}, \citenamefont {Stanley},\ and\ \citenamefont
  {Makse}}]{kitsak2010identification}%
  \BibitemOpen
  \bibfield  {author} {\bibinfo {author} {\bibfnamefont {M.}~\bibnamefont
  {Kitsak}}, \bibinfo {author} {\bibfnamefont {L.~K.}\ \bibnamefont {Gallos}},
  \bibinfo {author} {\bibfnamefont {S.}~\bibnamefont {Havlin}}, \bibinfo
  {author} {\bibfnamefont {F.}~\bibnamefont {Liljeros}}, \bibinfo {author}
  {\bibfnamefont {L.}~\bibnamefont {Muchnik}}, \bibinfo {author} {\bibfnamefont
  {H.~E.}\ \bibnamefont {Stanley}}, \ and\ \bibinfo {author} {\bibfnamefont
  {H.~A.}\ \bibnamefont {Makse}},\ }\href@noop {} {\bibfield  {journal}
  {\bibinfo  {journal} {Nature Physics}\ }\textbf {\bibinfo {volume} {6}},\
  \bibinfo {pages} {888} (\bibinfo {year} {2010})}\BibitemShut {NoStop}%
\bibitem [{\citenamefont {Daley}\ and\ \citenamefont
  {Kendall}(1965)}]{daley1965stochastic}%
  \BibitemOpen
  \bibfield  {author} {\bibinfo {author} {\bibfnamefont {D.}~\bibnamefont
  {Daley}}\ and\ \bibinfo {author} {\bibfnamefont {D.~G.}\ \bibnamefont
  {Kendall}},\ }\href@noop {} {\bibfield  {journal} {\bibinfo  {journal} {IMA
  Journal of Applied Mathematics}\ }\textbf {\bibinfo {volume} {1}},\ \bibinfo
  {pages} {42} (\bibinfo {year} {1965})}\BibitemShut {NoStop}%
\bibitem [{\citenamefont {Ferguson}\ \emph {et~al.}(2006)\citenamefont
  {Ferguson}, \citenamefont {Cummings}, \citenamefont {Fraser}, \citenamefont
  {Cajka}, \citenamefont {Cooley},\ and\ \citenamefont
  {Burke}}]{ferguson2006strategies}%
  \BibitemOpen
  \bibfield  {author} {\bibinfo {author} {\bibfnamefont {N.~M.}\ \bibnamefont
  {Ferguson}}, \bibinfo {author} {\bibfnamefont {D.~A.}\ \bibnamefont
  {Cummings}}, \bibinfo {author} {\bibfnamefont {C.}~\bibnamefont {Fraser}},
  \bibinfo {author} {\bibfnamefont {J.~C.}\ \bibnamefont {Cajka}}, \bibinfo
  {author} {\bibfnamefont {P.~C.}\ \bibnamefont {Cooley}}, \ and\ \bibinfo
  {author} {\bibfnamefont {D.~S.}\ \bibnamefont {Burke}},\ }\href@noop {}
  {\bibfield  {journal} {\bibinfo  {journal} {Nature}\ }\textbf {\bibinfo
  {volume} {442}},\ \bibinfo {pages} {448} (\bibinfo {year}
  {2006})}\BibitemShut {NoStop}%
\bibitem [{\citenamefont {Degli~Atti}\ \emph {et~al.}(2008)\citenamefont
  {Degli~Atti}, \citenamefont {Merler}, \citenamefont {Rizzo}, \citenamefont
  {Ajelli}, \citenamefont {Massari}, \citenamefont {Manfredi}, \citenamefont
  {Furlanello}, \citenamefont {Tomba},\ and\ \citenamefont
  {Iannelli}}]{degli2008mitigation}%
  \BibitemOpen
  \bibfield  {author} {\bibinfo {author} {\bibfnamefont {M.~L.~C.}\
  \bibnamefont {Degli~Atti}}, \bibinfo {author} {\bibfnamefont
  {S.}~\bibnamefont {Merler}}, \bibinfo {author} {\bibfnamefont
  {C.}~\bibnamefont {Rizzo}}, \bibinfo {author} {\bibfnamefont
  {M.}~\bibnamefont {Ajelli}}, \bibinfo {author} {\bibfnamefont
  {M.}~\bibnamefont {Massari}}, \bibinfo {author} {\bibfnamefont
  {P.}~\bibnamefont {Manfredi}}, \bibinfo {author} {\bibfnamefont
  {C.}~\bibnamefont {Furlanello}}, \bibinfo {author} {\bibfnamefont {G.~S.}\
  \bibnamefont {Tomba}}, \ and\ \bibinfo {author} {\bibfnamefont
  {M.}~\bibnamefont {Iannelli}},\ }\href@noop {} {\bibfield  {journal}
  {\bibinfo  {journal} {PLoS One}\ }\textbf {\bibinfo {volume} {3}},\ \bibinfo
  {pages} {e1790} (\bibinfo {year} {2008})}\BibitemShut {NoStop}%
\bibitem [{\citenamefont {Balcan}\ \emph {et~al.}(2010)\citenamefont {Balcan},
  \citenamefont {Gon{\c{c}}alves}, \citenamefont {Hu}, \citenamefont {Ramasco},
  \citenamefont {Colizza},\ and\ \citenamefont
  {Vespignani}}]{balcan2010modeling}%
  \BibitemOpen
  \bibfield  {author} {\bibinfo {author} {\bibfnamefont {D.}~\bibnamefont
  {Balcan}}, \bibinfo {author} {\bibfnamefont {B.}~\bibnamefont
  {Gon{\c{c}}alves}}, \bibinfo {author} {\bibfnamefont {H.}~\bibnamefont {Hu}},
  \bibinfo {author} {\bibfnamefont {J.~J.}\ \bibnamefont {Ramasco}}, \bibinfo
  {author} {\bibfnamefont {V.}~\bibnamefont {Colizza}}, \ and\ \bibinfo
  {author} {\bibfnamefont {A.}~\bibnamefont {Vespignani}},\ }\href@noop {}
  {\bibfield  {journal} {\bibinfo  {journal} {Journal of {C}omputational
  {S}cience}\ }\textbf {\bibinfo {volume} {1}},\ \bibinfo {pages} {132}
  (\bibinfo {year} {2010})}\BibitemShut {NoStop}%
\bibitem [{\citenamefont {Allen}(1994)}]{allen1994some}%
  \BibitemOpen
  \bibfield  {author} {\bibinfo {author} {\bibfnamefont {L.~J.}\ \bibnamefont
  {Allen}},\ }\href@noop {} {\bibfield  {journal} {\bibinfo  {journal}
  {Mathematical {B}iosciences}\ }\textbf {\bibinfo {volume} {124}},\ \bibinfo
  {pages} {83} (\bibinfo {year} {1994})}\BibitemShut {NoStop}%
\bibitem [{\citenamefont {Wang}\ \emph {et~al.}(2003)\citenamefont {Wang},
  \citenamefont {Chakrabarti}, \citenamefont {Wang},\ and\ \citenamefont
  {Faloutsos}}]{wang2003epidemic}%
  \BibitemOpen
  \bibfield  {author} {\bibinfo {author} {\bibfnamefont {Y.}~\bibnamefont
  {Wang}}, \bibinfo {author} {\bibfnamefont {D.}~\bibnamefont {Chakrabarti}},
  \bibinfo {author} {\bibfnamefont {C.}~\bibnamefont {Wang}}, \ and\ \bibinfo
  {author} {\bibfnamefont {C.}~\bibnamefont {Faloutsos}},\ }in\ \href@noop {}
  {\emph {\bibinfo {booktitle} {Reliable Distributed Systems, 2003.
  Proceedings. 22nd International Symposium on}}}\ (\bibinfo {organization}
  {IEEE},\ \bibinfo {year} {2003})\ pp.\ \bibinfo {pages} {25--34}\BibitemShut
  {NoStop}%
\bibitem [{\citenamefont {Zou}\ \emph {et~al.}(2007)\citenamefont {Zou},
  \citenamefont {Towsley},\ and\ \citenamefont {Gong}}]{zou2007modeling}%
  \BibitemOpen
  \bibfield  {author} {\bibinfo {author} {\bibfnamefont {C.~C.}\ \bibnamefont
  {Zou}}, \bibinfo {author} {\bibfnamefont {D.}~\bibnamefont {Towsley}}, \ and\
  \bibinfo {author} {\bibfnamefont {W.}~\bibnamefont {Gong}},\ }\href@noop {}
  {\bibfield  {journal} {\bibinfo  {journal} {Dependable and Secure Computing,
  IEEE Transactions on}\ }\textbf {\bibinfo {volume} {4}},\ \bibinfo {pages}
  {105} (\bibinfo {year} {2007})}\BibitemShut {NoStop}%
\bibitem [{\citenamefont {Chakrabarti}\ \emph {et~al.}(2008)\citenamefont
  {Chakrabarti}, \citenamefont {Wang}, \citenamefont {Wang}, \citenamefont
  {Leskovec},\ and\ \citenamefont {Faloutsos}}]{chakrabarti2008epidemic}%
  \BibitemOpen
  \bibfield  {author} {\bibinfo {author} {\bibfnamefont {D.}~\bibnamefont
  {Chakrabarti}}, \bibinfo {author} {\bibfnamefont {Y.}~\bibnamefont {Wang}},
  \bibinfo {author} {\bibfnamefont {C.}~\bibnamefont {Wang}}, \bibinfo {author}
  {\bibfnamefont {J.}~\bibnamefont {Leskovec}}, \ and\ \bibinfo {author}
  {\bibfnamefont {C.}~\bibnamefont {Faloutsos}},\ }\href@noop {} {\bibfield
  {journal} {\bibinfo  {journal} {ACM Transactions on Information and System
  Security (TISSEC)}\ }\textbf {\bibinfo {volume} {10}},\ \bibinfo {pages} {1}
  (\bibinfo {year} {2008})}\BibitemShut {NoStop}%
\bibitem [{\citenamefont {G{\'o}mez}\ \emph {et~al.}(2010)\citenamefont
  {G{\'o}mez}, \citenamefont {Arenas}, \citenamefont {Borge-Holthoefer},
  \citenamefont {Meloni},\ and\ \citenamefont {Moreno}}]{gomez2010discrete}%
  \BibitemOpen
  \bibfield  {author} {\bibinfo {author} {\bibfnamefont {S.}~\bibnamefont
  {G{\'o}mez}}, \bibinfo {author} {\bibfnamefont {A.}~\bibnamefont {Arenas}},
  \bibinfo {author} {\bibfnamefont {J.}~\bibnamefont {Borge-Holthoefer}},
  \bibinfo {author} {\bibfnamefont {S.}~\bibnamefont {Meloni}}, \ and\ \bibinfo
  {author} {\bibfnamefont {Y.}~\bibnamefont {Moreno}},\ }\href@noop {}
  {\bibfield  {journal} {\bibinfo  {journal} {EPL (Europhysics Letters)}\
  }\textbf {\bibinfo {volume} {89}},\ \bibinfo {pages} {38009} (\bibinfo {year}
  {2010})}\BibitemShut {NoStop}%
\bibitem [{\citenamefont {De~Domenico}\ \emph {et~al.}(2013)\citenamefont
  {De~Domenico}, \citenamefont {Lima}, \citenamefont {Mougel},\ and\
  \citenamefont {Musolesi}}]{de2013anatomy}%
  \BibitemOpen
  \bibfield  {author} {\bibinfo {author} {\bibfnamefont {M.}~\bibnamefont
  {De~Domenico}}, \bibinfo {author} {\bibfnamefont {A.}~\bibnamefont {Lima}},
  \bibinfo {author} {\bibfnamefont {P.}~\bibnamefont {Mougel}}, \ and\ \bibinfo
  {author} {\bibfnamefont {M.}~\bibnamefont {Musolesi}},\ }\href@noop {}
  {\bibfield  {journal} {\bibinfo  {journal} {Scientific {R}eports}\ }\textbf
  {\bibinfo {volume} {3}} (\bibinfo {year} {2013})}\BibitemShut {NoStop}%
\bibitem [{\citenamefont {Wei}\ \emph {et~al.}(2013)\citenamefont {Wei},
  \citenamefont {Valler}, \citenamefont {Prakash}, \citenamefont {Neamtiu},
  \citenamefont {Faloutsos},\ and\ \citenamefont
  {Faloutsos}}]{wei2013competing}%
  \BibitemOpen
  \bibfield  {author} {\bibinfo {author} {\bibfnamefont {X.}~\bibnamefont
  {Wei}}, \bibinfo {author} {\bibfnamefont {N.~C.}\ \bibnamefont {Valler}},
  \bibinfo {author} {\bibfnamefont {B.~A.}\ \bibnamefont {Prakash}}, \bibinfo
  {author} {\bibfnamefont {I.}~\bibnamefont {Neamtiu}}, \bibinfo {author}
  {\bibfnamefont {M.}~\bibnamefont {Faloutsos}}, \ and\ \bibinfo {author}
  {\bibfnamefont {C.}~\bibnamefont {Faloutsos}},\ }\href@noop {} {\bibfield
  {journal} {\bibinfo  {journal} {Selected Areas in Communications, IEEE
  Journal on}\ }\textbf {\bibinfo {volume} {31}},\ \bibinfo {pages} {1049}
  (\bibinfo {year} {2013})}\BibitemShut {NoStop}%
\bibitem [{\citenamefont {Granell}\ \emph {et~al.}(2013)\citenamefont
  {Granell}, \citenamefont {G{\'o}mez},\ and\ \citenamefont
  {Arenas}}]{granell2013dynamical}%
  \BibitemOpen
  \bibfield  {author} {\bibinfo {author} {\bibfnamefont {C.}~\bibnamefont
  {Granell}}, \bibinfo {author} {\bibfnamefont {S.}~\bibnamefont {G{\'o}mez}},
  \ and\ \bibinfo {author} {\bibfnamefont {A.}~\bibnamefont {Arenas}},\
  }\href@noop {} {\bibfield  {journal} {\bibinfo  {journal} {Physical review
  letters}\ }\textbf {\bibinfo {volume} {111}},\ \bibinfo {pages} {128701}
  (\bibinfo {year} {2013})}\BibitemShut {NoStop}%
\bibitem [{\citenamefont {Cozzo}\ \emph {et~al.}(2013)\citenamefont {Cozzo},
  \citenamefont {Banos}, \citenamefont {Meloni},\ and\ \citenamefont
  {Moreno}}]{cozzo2013contact}%
  \BibitemOpen
  \bibfield  {author} {\bibinfo {author} {\bibfnamefont {E.}~\bibnamefont
  {Cozzo}}, \bibinfo {author} {\bibfnamefont {R.~A.}\ \bibnamefont {Banos}},
  \bibinfo {author} {\bibfnamefont {S.}~\bibnamefont {Meloni}}, \ and\ \bibinfo
  {author} {\bibfnamefont {Y.}~\bibnamefont {Moreno}},\ }\href@noop {}
  {\bibfield  {journal} {\bibinfo  {journal} {Physical Review E}\ }\textbf
  {\bibinfo {volume} {88}},\ \bibinfo {pages} {050801} (\bibinfo {year}
  {2013})}\BibitemShut {NoStop}%
\bibitem [{\citenamefont {Valdano}\ \emph {et~al.}(2015)\citenamefont
  {Valdano}, \citenamefont {Ferreri}, \citenamefont {Poletto},\ and\
  \citenamefont {Colizza}}]{valdano2015analytical}%
  \BibitemOpen
  \bibfield  {author} {\bibinfo {author} {\bibfnamefont {E.}~\bibnamefont
  {Valdano}}, \bibinfo {author} {\bibfnamefont {L.}~\bibnamefont {Ferreri}},
  \bibinfo {author} {\bibfnamefont {C.}~\bibnamefont {Poletto}}, \ and\
  \bibinfo {author} {\bibfnamefont {V.}~\bibnamefont {Colizza}},\ }\href@noop
  {} {\bibfield  {journal} {\bibinfo  {journal} {Physical Review X}\ }\textbf
  {\bibinfo {volume} {5}},\ \bibinfo {pages} {021005} (\bibinfo {year}
  {2015})}\BibitemShut {NoStop}%
\bibitem [{\citenamefont {G{\'o}mez-Garde{\~n}es}\ \emph
  {et~al.}(2016)\citenamefont {G{\'o}mez-Garde{\~n}es}, \citenamefont {Lotero},
  \citenamefont {Taraskin},\ and\ \citenamefont
  {P{\'e}rez-Reche}}]{gomez2016explosive}%
  \BibitemOpen
  \bibfield  {author} {\bibinfo {author} {\bibfnamefont {J.}~\bibnamefont
  {G{\'o}mez-Garde{\~n}es}}, \bibinfo {author} {\bibfnamefont {L.}~\bibnamefont
  {Lotero}}, \bibinfo {author} {\bibfnamefont {S.}~\bibnamefont {Taraskin}}, \
  and\ \bibinfo {author} {\bibfnamefont {F.}~\bibnamefont {P{\'e}rez-Reche}},\
  }\href@noop {} {\bibfield  {journal} {\bibinfo  {journal} {Scientific
  {R}eports}\ }\textbf {\bibinfo {volume} {6}} (\bibinfo {year}
  {2016})}\BibitemShut {NoStop}%
\bibitem [{\citenamefont {G{\'o}mez}\ \emph {et~al.}(2011)\citenamefont
  {G{\'o}mez}, \citenamefont {G{\'o}mez-Gardenes}, \citenamefont {Moreno},\
  and\ \citenamefont {Arenas}}]{gomez2011nonperturbative}%
  \BibitemOpen
  \bibfield  {author} {\bibinfo {author} {\bibfnamefont {S.}~\bibnamefont
  {G{\'o}mez}}, \bibinfo {author} {\bibfnamefont {J.}~\bibnamefont
  {G{\'o}mez-Gardenes}}, \bibinfo {author} {\bibfnamefont {Y.}~\bibnamefont
  {Moreno}}, \ and\ \bibinfo {author} {\bibfnamefont {A.}~\bibnamefont
  {Arenas}},\ }\href@noop {} {\bibfield  {journal} {\bibinfo  {journal}
  {Physical Review E}\ }\textbf {\bibinfo {volume} {84}},\ \bibinfo {pages}
  {036105} (\bibinfo {year} {2011})}\BibitemShut {NoStop}%
\bibitem [{\citenamefont {Barth{\'e}lemy}\ \emph {et~al.}(2004)\citenamefont
  {Barth{\'e}lemy}, \citenamefont {Barrat}, \citenamefont {Pastor-Satorras},\
  and\ \citenamefont {Vespignani}}]{barthelemy2004velocity}%
  \BibitemOpen
  \bibfield  {author} {\bibinfo {author} {\bibfnamefont {M.}~\bibnamefont
  {Barth{\'e}lemy}}, \bibinfo {author} {\bibfnamefont {A.}~\bibnamefont
  {Barrat}}, \bibinfo {author} {\bibfnamefont {R.}~\bibnamefont
  {Pastor-Satorras}}, \ and\ \bibinfo {author} {\bibfnamefont {A.}~\bibnamefont
  {Vespignani}},\ }\href@noop {} {\bibfield  {journal} {\bibinfo  {journal}
  {Physical Review Letters}\ }\textbf {\bibinfo {volume} {92}},\ \bibinfo
  {pages} {178701} (\bibinfo {year} {2004})}\BibitemShut {NoStop}%
\bibitem [{\citenamefont {Bailey}(1975)}]{bailey1975mathematical}%
  \BibitemOpen
  \bibfield  {author} {\bibinfo {author} {\bibfnamefont {N.~T.}\ \bibnamefont
  {Bailey}},\ }\href@noop {} {\emph {\bibinfo {title} {The {M}athematical
  {T}heory of {I}nfectious {D}iseases and its {A}pplications}}}\ (\bibinfo
  {publisher} {Charles Griffin \& Company Ltd},\ \bibinfo {year}
  {1975})\BibitemShut {NoStop}%
\bibitem [{\citenamefont {Liggett}(1999)}]{liggett1999}%
  \BibitemOpen
  \bibfield  {author} {\bibinfo {author} {\bibfnamefont {T.~M.}\ \bibnamefont
  {Liggett}},\ }\href@noop {} {\emph {\bibinfo {title} {{S}tochastic
  {I}nteracting {S}ystems: {C}ontact, {V}oter and {E}xclusion processes}}},\
  Vol.\ \bibinfo {volume} {324}\ (\bibinfo  {publisher} {Springer},\ \bibinfo
  {year} {1999})\BibitemShut {NoStop}%
\bibitem [{\citenamefont {Gardiner}(1985)}]{gardiner1985handbook}%
  \BibitemOpen
  \bibfield  {author} {\bibinfo {author} {\bibfnamefont {C.~W.}\ \bibnamefont
  {Gardiner}},\ }\href@noop {} {\bibfield  {journal} {\bibinfo  {journal}
  {Springer Series in Synergetics}\ } (\bibinfo {year} {1985})}\BibitemShut
  {NoStop}%
\bibitem [{\citenamefont {Liggett}(1985)}]{liggett1985particle}%
  \BibitemOpen
  \bibfield  {author} {\bibinfo {author} {\bibfnamefont {T.~M.}\ \bibnamefont
  {Liggett}},\ }\href@noop {} {\emph {\bibinfo {title} {Interacting {P}article
  {S}ystems}}}\ (\bibinfo  {publisher} {Springer},\ \bibinfo {year}
  {1985})\BibitemShut {NoStop}%
\bibitem [{\citenamefont {Gillespie}(1977)}]{gillespie1977}%
  \BibitemOpen
  \bibfield  {author} {\bibinfo {author} {\bibfnamefont {D.~T.}\ \bibnamefont
  {Gillespie}},\ }\href@noop {} {\bibfield  {journal} {\bibinfo  {journal} {The
  Journal of Physical Chemistry}\ }\textbf {\bibinfo {volume} {81}},\ \bibinfo
  {pages} {2340} (\bibinfo {year} {1977})}\BibitemShut {NoStop}%
\bibitem [{\citenamefont {Ferreira}\ \emph {et~al.}(2012)\citenamefont
  {Ferreira}, \citenamefont {Castellano},\ and\ \citenamefont
  {Pastor-Satorras}}]{ferreira2012epidemic}%
  \BibitemOpen
  \bibfield  {author} {\bibinfo {author} {\bibfnamefont {S.~C.}\ \bibnamefont
  {Ferreira}}, \bibinfo {author} {\bibfnamefont {C.}~\bibnamefont
  {Castellano}}, \ and\ \bibinfo {author} {\bibfnamefont {R.}~\bibnamefont
  {Pastor-Satorras}},\ }\href@noop {} {\bibfield  {journal} {\bibinfo
  {journal} {Physical Review E}\ }\textbf {\bibinfo {volume} {86}},\ \bibinfo
  {pages} {041125} (\bibinfo {year} {2012})}\BibitemShut {NoStop}%
\bibitem [{\citenamefont {Vestergaard}\ and\ \citenamefont
  {G{\'e}nois}(2015)}]{vestergaard2015temporal}%
  \BibitemOpen
  \bibfield  {author} {\bibinfo {author} {\bibfnamefont {C.~L.}\ \bibnamefont
  {Vestergaard}}\ and\ \bibinfo {author} {\bibfnamefont {M.}~\bibnamefont
  {G{\'e}nois}},\ }\href@noop {} {\bibfield  {journal} {\bibinfo  {journal}
  {PLoS Comput Biol}\ }\textbf {\bibinfo {volume} {11}},\ \bibinfo {pages}
  {e1004579} (\bibinfo {year} {2015})}\BibitemShut {NoStop}%
\bibitem [{\citenamefont {Voter}(2007)}]{voter2007introduction}%
  \BibitemOpen
  \bibfield  {author} {\bibinfo {author} {\bibfnamefont {A.~F.}\ \bibnamefont
  {Voter}},\ }in\ \href@noop {} {\emph {\bibinfo {booktitle} {Radiation Effects
  in Solids}}}\ (\bibinfo  {publisher} {Springer},\ \bibinfo {year} {2007})\
  pp.\ \bibinfo {pages} {1--23}\BibitemShut {NoStop}%
\bibitem [{\citenamefont {Neal}(2003)}]{neal2003slice}%
  \BibitemOpen
  \bibfield  {author} {\bibinfo {author} {\bibfnamefont {R.~M.}\ \bibnamefont
  {Neal}},\ }\href@noop {} {\bibfield  {journal} {\bibinfo  {journal} {Annals
  of {S}tatistics}\ ,\ \bibinfo {pages} {705}} (\bibinfo {year}
  {2003})}\BibitemShut {NoStop}%
\bibitem [{web(2016)}]{website:SIScode}%
  \BibitemOpen
  \href@noop {} {}\bibinfo {howpublished}
  {\url{https://github.com/peterfennell/SIS-simulations}} (\bibinfo {year}
  {2016})\BibitemShut {NoStop}%
\bibitem [{Note1()}]{Note1}%
  \BibitemOpen
  \bibinfo {note} {We refer to the metastable state as the state when the
  expected fraction of infected nodes has reached a plateau which very slowly
  decreases towards 0.}\BibitemShut {Stop}%
\bibitem [{\citenamefont {Castellano}\ and\ \citenamefont
  {Pastor-Satorras}(2010)}]{castellano2010thresholds}%
  \BibitemOpen
  \bibfield  {author} {\bibinfo {author} {\bibfnamefont {C.}~\bibnamefont
  {Castellano}}\ and\ \bibinfo {author} {\bibfnamefont {R.}~\bibnamefont
  {Pastor-Satorras}},\ }\href@noop {} {\bibfield  {journal} {\bibinfo
  {journal} {Physical Review Letters}\ }\textbf {\bibinfo {volume} {105}},\
  \bibinfo {pages} {218701} (\bibinfo {year} {2010})}\BibitemShut {NoStop}%
\bibitem [{\citenamefont {Goltsev}\ \emph {et~al.}(2012)\citenamefont
  {Goltsev}, \citenamefont {Dorogovtsev}, \citenamefont {Oliveira},\ and\
  \citenamefont {Mendes}}]{goltsev2012localization}%
  \BibitemOpen
  \bibfield  {author} {\bibinfo {author} {\bibfnamefont {A.~V.}\ \bibnamefont
  {Goltsev}}, \bibinfo {author} {\bibfnamefont {S.~N.}\ \bibnamefont
  {Dorogovtsev}}, \bibinfo {author} {\bibfnamefont {J.}~\bibnamefont
  {Oliveira}}, \ and\ \bibinfo {author} {\bibfnamefont {J.~F.}\ \bibnamefont
  {Mendes}},\ }\href@noop {} {\bibfield  {journal} {\bibinfo  {journal}
  {Physical Review Letters}\ }\textbf {\bibinfo {volume} {109}},\ \bibinfo
  {pages} {128702} (\bibinfo {year} {2012})}\BibitemShut {NoStop}%
\bibitem [{\citenamefont {Van~Mieghem}(2012)}]{van2012epidemic}%
  \BibitemOpen
  \bibfield  {author} {\bibinfo {author} {\bibfnamefont {P.}~\bibnamefont
  {Van~Mieghem}},\ }\href@noop {} {\bibfield  {journal} {\bibinfo  {journal}
  {EPL (Europhysics Letters)}\ }\textbf {\bibinfo {volume} {97}},\ \bibinfo
  {pages} {48004} (\bibinfo {year} {2012})}\BibitemShut {NoStop}%
\bibitem [{\citenamefont {Bogu{\~n}{\'a}}\ \emph {et~al.}(2013)\citenamefont
  {Bogu{\~n}{\'a}}, \citenamefont {Castellano},\ and\ \citenamefont
  {Pastor-Satorras}}]{boguna2013nature}%
  \BibitemOpen
  \bibfield  {author} {\bibinfo {author} {\bibfnamefont {M.}~\bibnamefont
  {Bogu{\~n}{\'a}}}, \bibinfo {author} {\bibfnamefont {C.}~\bibnamefont
  {Castellano}}, \ and\ \bibinfo {author} {\bibfnamefont {R.}~\bibnamefont
  {Pastor-Satorras}},\ }\href@noop {} {\bibfield  {journal} {\bibinfo
  {journal} {Physical Review Letters}\ }\textbf {\bibinfo {volume} {111}},\
  \bibinfo {pages} {068701} (\bibinfo {year} {2013})}\BibitemShut {NoStop}%
\bibitem [{\citenamefont {Mata}\ and\ \citenamefont
  {Ferreira}(2015)}]{mata2015multiple}%
  \BibitemOpen
  \bibfield  {author} {\bibinfo {author} {\bibfnamefont {A.~S.}\ \bibnamefont
  {Mata}}\ and\ \bibinfo {author} {\bibfnamefont {S.~C.}\ \bibnamefont
  {Ferreira}},\ }\href@noop {} {\bibfield  {journal} {\bibinfo  {journal}
  {Physical Review E}\ }\textbf {\bibinfo {volume} {91}},\ \bibinfo {pages}
  {012816} (\bibinfo {year} {2015})}\BibitemShut {NoStop}%
\bibitem [{\citenamefont {Kephart}\ and\ \citenamefont
  {White}(1991)}]{kephart1991directed}%
  \BibitemOpen
  \bibfield  {author} {\bibinfo {author} {\bibfnamefont {J.~O.}\ \bibnamefont
  {Kephart}}\ and\ \bibinfo {author} {\bibfnamefont {S.~R.}\ \bibnamefont
  {White}},\ }in\ \href@noop {} {\emph {\bibinfo {booktitle} {Proceedings of
  the 1991 IEEE Computer Society Symposium on Research in Security and
  Privacy}}}\ (\bibinfo {organization} {IEEE},\ \bibinfo {year} {1991})\ pp.\
  \bibinfo {pages} {343--359}\BibitemShut {NoStop}%
\bibitem [{\citenamefont {Van~Mieghem}\ \emph {et~al.}(2009)\citenamefont
  {Van~Mieghem}, \citenamefont {Omic},\ and\ \citenamefont
  {Kooij}}]{van2009virus}%
  \BibitemOpen
  \bibfield  {author} {\bibinfo {author} {\bibfnamefont {P.}~\bibnamefont
  {Van~Mieghem}}, \bibinfo {author} {\bibfnamefont {J.}~\bibnamefont {Omic}}, \
  and\ \bibinfo {author} {\bibfnamefont {R.}~\bibnamefont {Kooij}},\
  }\href@noop {} {\bibfield  {journal} {\bibinfo  {journal} {Networking,
  IEEE/ACM Transactions on}\ }\textbf {\bibinfo {volume} {17}},\ \bibinfo
  {pages} {1} (\bibinfo {year} {2009})}\BibitemShut {NoStop}%
\end{thebibliography}%

\end{document}